\documentclass[12pt,onecolumn,journal,draftcls]{IEEEtran}
\usepackage{epsf,psfrag,amssymb,amsfonts,amsmath,graphicx,cite,subfigure,geometry}
\def\defeq{{\stackrel{\Delta}{=}}}

\def\Sbf{{\bf S}}
\def\Kbf{{\bf K}}
\def\Ibf{{\bf I}}
\def\Ybf{{\bf Y}}
\def\kbf{{\bf k}}
\def\sbf{{\bf s}}
\def\Ac{{\cal A}}
\def\Ec{{\cal E}}
\def\Fc{{\cal F}}
\def\Hc{{\cal H}}
\def\Nc{{\cal N}}

\def\Tc{{\cal T}}
\def\alphabf{{\mbox{\boldmath$\alpha$\unboldmath}}}
\def\betabf{{\mbox{\boldmath$\beta$\unboldmath}}}
\def\thetabf{{\mbox{\boldmath$\theta$\unboldmath}}}
\def\thetabfs{{\mbox{\footnotesize \boldmath$\theta$\unboldmath}}}

\def\Thetabf{{\mbox{\boldmath$\Theta$\unboldmath}}}
\def\Lambdabf{{\mbox{\boldmath$\Lambda$\unboldmath}}}
\def\Phibf{{\mbox{\boldmath$\Phi$\unboldmath}}}
\def\Thetabfs{{\mbox{\footnotesize \boldmath$\Theta$\unboldmath}}}

\def\eg{{\it e.g.,\ \/}}
\def\ie{{\it i.e.,\ \/}}
\renewcommand{\QED}{\QEDopen\QEDopen\QEDopen}
\def\scalefig#1{\epsfxsize #1\textwidth}
\newtheorem{theorem}{Theorem}
\newtheorem{lemma}{Lemma}
\newtheorem{proposition}{Proposition}

\begin{document}

\title{\bf \Large Joint Design and Separation Principle for
Opportunistic\\[-0.2em]
Spectrum Access in the Presence of Sensing Errors\thanks{This work
was supported in part by the Army Research Laboratory CTA on
Communication and Networks under Grant DAAD19-01-2-0011 and by the
National Science Foundation under Grants CNS-0627090 and
ECS-0622200. Part of this work was presented in the {\em 39th Annual
Asilomar Conference on Signal, Systems, and Computers.}, Oct. --
Nov., 2006 and submitted to the {\em IEEE Workshop on Signal
Processing Advances in Wireless Communications}, 2007.}}
\author{\vspace{-0.5em}Yunxia Chen, Qing Zhao$^*$, and
Ananthram Swami\thanks{Yunxia Chen and Qing Zhao are with the
Department of Electrical and Computer Engineering, University of
California, Davis, CA 95616. Emails:
\{yxchen,qzhao\}@ece.ucdavis.edu.  Ananthram Swami is with the Army
Research Laboratory, Adelphi, MD 20783. Email: aswami@arl.army.mil.}
\thanks{$*$ Corresponding author. Phone: 1-530-752-7390. Fax:
1-530-752-8428.}}

\markboth{Submitted to {\em IEEE Transactions on Information Theory}, Feb. 2007.}{Chen, Zhao, and Swami} \maketitle%
\thispagestyle{empty}

\vspace{-4.5em}

\begin{abstract}

\vspace{-.5em} We address the design of opportunistic spectrum
access (OSA) strategies that allow secondary users to independently
search for and exploit instantaneous spectrum availability. The
design objective is to maximize the throughput of secondary users
while limiting the probability of colliding with primary users.
Integrated in the joint design are three basic components: a
spectrum sensor at the physical (PHY) layer that identifies spectrum
opportunities, a sensing strategy at the medium access control (MAC)
layer that determines which channels in the spectrum to sense, and
an access strategy, also at the MAC layer, that decides whether to
access based on sensing outcomes that are subject to errors.

We formulate the joint PHY-MAC design of OSA as a constrained
partially observable Markov decision process (POMDP). Constrained
POMDPs generally require randomized policies to achieve optimality,
which are often intractable. By exploiting the rich structure of the
underlying problem, we establish a separation principle for the
joint design of OSA. Specifically, the optimal joint design can be
carried out in two steps: first to choose the spectrum sensor and
the access strategy to maximize the instantaneous throughput under a
collision constraint, and then to choose the sensing strategy to
maximize the overall throughput. This separation principle reveals
the optimality of myopic policies for the design of the spectrum
sensor and the access strategy, leading to closed-form optimal
solutions. Furthermore, decoupling the design of the sensing
strategy from that of the spectrum sensor and the access strategy,
the separation principle reduces the constrained POMDP to an
unconstrained one, which admits deterministic optimal policies.
Numerical examples are provided to study the design tradeoffs, the
interaction between the PHY layer spectrum sensor and the MAC layer
sensing and access strategies, and the robustness of the ensuing
design to model mismatch.
\end{abstract}

\vspace{-.5em}
\begin{keywords}
\vspace{-.5em} Opportunistic spectrum access, partially observable
Markov decision process.
\end{keywords}

\newpage
\section{Introduction}\label{sec:intro}
The exponential growth in wireless services and the physical limit
on usable radio frequencies have motivated various dynmaic spectrum
sharing strategies, among which is opportunistic spectrum access
(OSA). OSA, first envisioned by Mitola \cite{Mitola:99MoMuC} under
the term ``spectrum pooling'' and then investigated by the DARPA XG
program \cite{DARPA:XG}, has recently received increasing attention
due to its potential for improving spectrum efficiency
\cite{DySPAN05,CrownCom06}. The basic idea of OSA is to allow
secondary users to search for, identify, and exploit instantaneous
spectrum opportunities while limiting the level of interference
perceived by primary users (or licensees).

In this paper, we address the design of OSA strategies for secondary
users overlaying a slotted primary network. Integrated in the OSA
design are three basic components: 1) a spectrum sensor at the
physical (PHY) layer that identifies instantaneous spectrum
opportunities; 2) a spectrum sensing strategy at the  medium access
control (MAC) layer that specifies which channels in the spectrum to
sense in each slot; and 3) a spectrum access strategy, also at the
MAC layer, that determines whether to access the chosen channels
based on imperfect sensing outcomes. The design objective is to
maximize the throughput of secondary users under the constraint that
the probability of collision perceived by any primary user is below
a pre-determined threshold.

\subsection{Fundamental Design Tradeoffs}
We provide first an intuitive understanding of the fundamental
tradeoffs in the joint design of the three basic components.

\vspace{0.5em}

\begin{figure}[]
\centerline{\scalefig{0.7}\epsfbox{ 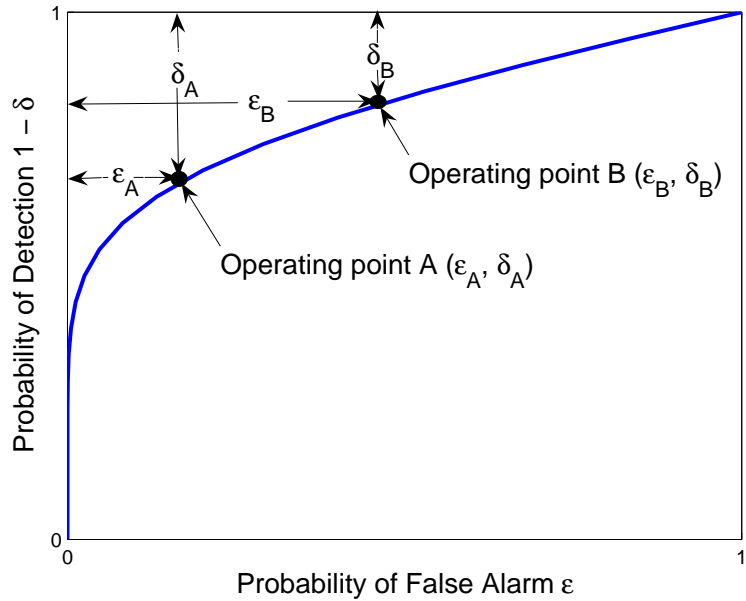}} \caption{The
ROC of an energy detector. Each point on the ROC curve corresponds
to a sensor operating characteristic resulting from different
detection threshold of the energy detector. ($\epsilon$: probability
of false alarm, $\delta$: probability of miss detection.)}
\label{fig:ROCexample}
\end{figure}

\noindent{\em Spectrum Sensor: False Alarm vs. Miss Detection}~~~
The spectrum sensor of a secondary user identifies spectrum
opportunities by detecting the presence of primary signals, \ie by
performing a binary hypothesis test. With noise and fading, sensing
errors are inevitable: false alarms occur when idle channels are
detected as busy, and miss detections occur when busy channels are
detected as idle. In the event of a false alarm, a spectrum
opportunity is overlooked by the sensor, and eventually wasted if
the access strategy trusts the sensing outcome. On the other hand,
miss detections may lead to collisions with primary users. The
tradeoff between false alarm and miss detection is captured by the
receiver operating characteristic (ROC) of the spectrum sensor,
which relates the probability of detection (PD) and the probability
of false alarm (PFA) (see an example in Fig.~\ref{fig:ROCexample}
where we consider an energy detector). The design of the spectrum
sensor and the choice of the sensor operating point are thus
important issues and should be addressed by considering the impact
of sensing errors on the MAC layer performance in terms of
throughput and collision probability. In particular, we are
interested in the fundamental question that which criterion should
be adopted in the design of the spectrum sensor, the Bayes or the
Neyman-Pearson (NP). If the former, how do we choose the risks? If
the latter, how should we set the constraint on the PFA?

\vspace{0.5em}

\noindent{\em Sensing Strategy: Gaining Immediate Access vs. Gaining
Information for Future Use}~~~ Due to hardware limitations and the
energy cost of spectrum monitoring, a secondary user may not be able
to sense all the channels in the spectrum simultaneously. A sensing
strategy is thus needed for intelligent channel selection to track
the rapidly varying spectrum opportunities. The purpose of a sensing
strategy is twofold: to find idle channels for immediate access and
to gain statistical information on the spectrum occupancy for better
opportunity tracking in the future. The optimal sensing strategy
should thus strike a balance between these two often conflicting
objectives.

\vspace{0.5em}

\noindent{\em Access Strategy: Aggressive vs. Conservative}~~~ Based
on the imperfect sensing outcomes given by the spectrum sensor, the
secondary user needs to decide whether to access. An aggressive
access strategy may lead to excessive collisions with primary users
while a conservative one may result in throughput degradation due to
overlooked opportunities. Whether to adopt an aggressive or a
conservative access strategy depends on the operating characteristic
(false alarm vs. miss detection) of the spectrum sensor and the
collision constraint at the MAC layer. Hence, a joint design of the
PHY layer spectrum sensor and the MAC layer access strategy is
necessary for optimality.

\subsection{Main Results}
By modeling primary users' spectrum occupancy as a Markov process,
we establish a decision-theoretic framework for the optimal joint
design of OSA based on the theory of partially observable Markov
decision processes (POMDPs). This framework captures the fundamental
design tradeoffs discussed above. Within this framework, the optimal
OSA strategy is given by the optimal policy of a constrained POMDP.

While powerful in problem modeling, POMDP suffers from the curse of
dimensionality and does not easily lend itself to tractable
solutions. Constraints on a POMDP further complicates the problem,
often demanding {\em randomized} policies to achieve optimality. Our
goal is to develop structural results that lead to simple yet
optimal solutions and shed light on the interaction between the PHY
and the MAC layers of OSA networks.

\vspace{0.5em}

{\em Single-Channel Sensing}~~~ We focus first on the case where the
secondary user can sense and access one channel in each slot (\eg in
the case of single carrier communications). We establish a
separation principle for the optimal joint design of OSA. We show
that the joint design can be carried out in two steps without losing
optimality: first to choose a spectrum sensor and an access strategy
that maximize the instantaneous throughput (\ie the expected number
of bits that can be delivered in the current slot) under the
collision constraint, and then to choose a sensing strategy to
optimize the overall throughput. As stated below, the significance
of this separation principle is twofold.

\begin{itemize}
\item The separation principle reveals the optimality of myopic
policies for the design of the spectrum sensor and the access
strategy. Myopic policies that aim solely at maximizing the
immediate reward ignore the impact of the current actions on the
future reward. Hence, obtaining myopic policies becomes a static
optimization problem instead of a sequential decision-making
problem. While myopic policies are rarely optimal for a general
POMDP, we show that the rich structure of the problem at hand
renders an exception. As a consequence, we are able to obtain an
explicit design of the optimum spectrum sensor and a closed-form
optimal access strategy. Moreover, this closed-form optimal design
allows us to characterize quantitatively the interaction between the
PHY layer spectrum sensor and the MAC layer access strategy.

\item The separation principle decouples the design of the sensing
strategy from that of the spectrum sensor and the access strategy.
More importantly, the design of the sensing strategy is reduced to
an {\em unconstrained} POMDP, which admits {\em deterministic}
optimal policies. Unconstrained POMDPs have been well studied, and
existing algorithms can be readily applied \cite{Sondik:71PhD,
Smallwood&Sondik:73OR, Monahan:82MS, Cheng:88PhD}.
\end{itemize}

We also provide simulation examples to study design tradeoffs. We
will see that miss detections are more harmful to the throughput of
the secondary user than false alarms. The tradeoff study between the
spectrum sensing time and the data transmission time indicates that
the spectrum sensor should take fewer channel measurements as the
maximum allowable probability of collision increases. In other
words, when the collision constraint is less restrictive, the
secondary user can spend less time in sensing, leaving more time in
a slot for data transmission. Robustness studies show that the
throughput loss due to inaccuracies in the assumed Markovian model
parameters is small, and more importantly, the probability of
collision perceived by the primary network is not affected by model
mismatch.

\vspace{0.5em}

{\em Multi-Channel Sensing}~~~ We then consider the scenario where
the secondary user can sense and access multiple channels
simultaneously in each slot. We show that the separation principle
still holds if the spectrum sensor and the access strategy are
designed independently across channels. We note that such
independent design is suboptimal since it ignores the potential
correlation among channel occupancies. We thus propose two heuristic
approaches to exploit channel correlation, one at the PHY layer and
the other at the MAC layer. Simulation results show that exploiting
channel correlation at the PHY layer is more effective than at the
MAC layer.

We also find that the performance of the PHY layer spectrum sensor
can improve over time by incorporating the MAC layer sensing and
access decisions. Such MAC layer decisions provide information on
the evolution of the primary users' spectrum occupancy, from which
the {\em a priori} probabilities of the hypotheses employed by the
spectrum sensor can be learned. This finding, along with the
quantitative characterization of the impact of the spectrum sensor
on the access strategy, illustrates the two-way interaction between
the PHY and the MAC layers: the necessity of incorporating the
sensor operating characteristics into the MAC design and the benefit
of exploiting the MAC layer information in the PHY design.

\subsection{Related Work}
Two types of spectrum opportunities have been considered in the
literature: spatial and temporal. A majority of existing work on OSA
focuses on exploiting spatial spectrum opportunities that are static
or slowly varying in time (see \cite{Zheng&Peng:05ICC,
Wang&Liu:05VTC, Steenstrup:05DySpan} and references therein). A
typical example application is the reuse of locally unused TV
broadcast bands. In this context, due to the slow temporal variation
of spectrum occupancy, realtime opportunity identification is not as
critical a component as in applications that exploit temporal
spectrum opportunities, and the existing work often assumes perfect
knowledge of spectrum opportunities in the whole spectrum at any
location.

The exploitation of temporal spectrum opportunities resulting from
the bursty traffic of primary users is addressed in
\cite{Papadimitratos&etal:05CMag, Zhao&etal:05DySPAN,
Chen&etal:06MILCOM, Zhao&etal:07WCNC} under the assumption of
perfect sensing. In \cite{Papadimitratos&etal:05CMag}, MAC protocols
are proposed for an ad hoc secondary network overlaying a GSM
cellular network. It is assumed that the secondary transmitter and
receiver exchange information on which channel to use through a
commonly agreed control channel. Different from this work, optimal
distributed MAC protocols developed in \cite{Zhao&etal:05DySPAN} can
synchronize the hopping patterns of the secondary transmitter and
receiver without the aid of additional control channels. More
recently, the design of optimal spectrum sensing and access
strategies in a fading environment is addressed under an energy
constraint in \cite{Chen&etal:06MILCOM}. In \cite{Zhao&etal:07WCNC},
access strategies for a slotted secondary user searching for
opportunities in an un-slotted primary network is considered, where
a round-robin single-channel sensing scheme is used. Modeling of
spectrum occupancy has been addressed in
\cite{Geirhofer&etal:06TAPAS}. Measurements obtained from spectrum
monitoring test-beds demonstrate the Makovian transition between
busy and idle channel states in wireless LAN.

Although the issue of spectrum sensing errors has been investigated
at the PHY layer \cite{Sahai&etal:04Allerton,
Cabric&etal:04Asilomar, Challapali&etal:04,
Wild&Ramchandran:05DySPAN, Ghasemi&Sousa:05DySPAN}, cognitive MAC
design in the presence of sensing errors has received little
attention. To the best of our knowledge, \cite{Zhao&etal:06JSAC} is
the first work that integrates the operating characteristic of the
spectrum sensor at the PHY layer with the MAC design. A heuristic
approach to the joint PHY-MAC design of OSA is proposed in
\cite{Zhao&etal:06JSAC}. In this paper, we establish a
decision-theoretic framework within which the optimal joint design
of OSA in the presence of sensing errors can be systematically
addressed and the interaction between the PHY and the MAC layers can
be quantitatively characterized. Interestingly, the separation
principle developed in this paper reveals that the heuristic
approach proposed in \cite{Zhao&etal:06JSAC} is optimal.

For an overview on challenges and recent developments in OSA,
readers are referred to \cite{Zhao&Sadler:07SPMag}.

\subsection{Organization}
This paper is organized as follows. Section II describes the network
model and the basic operations performed by a secondary user to
exploit spectrum opportunities. In Section III, we introduce the
three basic components of OSA and formulate their joint design as a
constrained POMDP. In Section IV, we establish the separation
principle for the optimal joint design of OSA with single-channel
sensing. Section V extends the separation principle to multi-channel
sensing scenarios. Section VI concludes this paper.


\section{Network Model}\label{sec:model}
Consider a spectrum that consists of $N$ channels (\eg different
frequency bands or tones in an OFDM system), each with bandwidth
$B_n$ ($n=1,\cdots,N$). These $N$ channels are licensed to a slotted
primary network. We model the spectrum occupancy as a discrete-time
homogenous Markov process with $2^N$ states. Specifically, let
$S_n(t) \in \{0\mbox{ (busy), }1\mbox{ (idle)}\}$ denote the
occupancy of channel $n$ in slot $t$. The spectrum occupancy state
(SOS) $\Sbf(t)\,\defeq\,[S_1(t), \ldots, S_N(t)]$ follows a discrete
Markov process with finite state space $\mathbb{S} \,\defeq\,
\{0,1\}^N$. The transition probabilities are denoted as
$\{P_{\sbf, \sbf'}\}_{ \substack{\sbf\in\mathbb{S} \\
\sbf'\in\mathbb{S}}}$, where $P_{\sbf, \sbf'}\,\defeq\, \Pr\{\Sbf(t)
= \sbf' \,|\, \Sbf(t-1) = \sbf\}$ is the probability that the SOS
transits from $\sbf \in \mathbb{S} $ to $\sbf'\in\mathbb{S}$ at the
beginning of slot $t$. Note that the transition probabilities are
determined by the dynamics of the primary traffic. We assume that
they are known and remain unchanged in $T$ slots.

\begin{figure}[]
\centerline{
\begin{psfrags} ~~~~~~~\psfrag{O}[l]{\footnotesize Spectrum
Sensing} \psfrag{D}[l]{\footnotesize Data Transmission}
\psfrag{A}[l]{\footnotesize Acknowledgement} \psfrag{s}[c]{ Slot}
\scalefig{0.5}\epsfbox{ 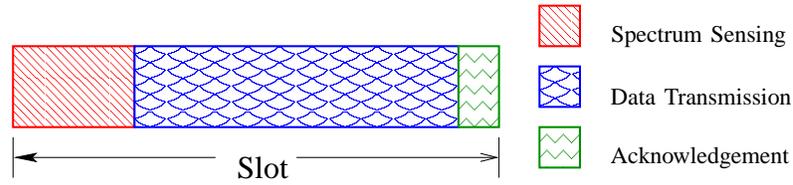}
\end{psfrags}~~~~~~~~~~~~~~~~~}
\caption{The slot structure.} \label{fig:slot}
\end{figure}

We consider a secondary ad hoc network whose users independently and
selfishly exploit instantaneous spectrum opportunities in these $N$
channels\footnote{We assume that the inter-channel interference is
negligible. Thus, a secondary user transmitting over an idle channel
does not interfere with primary users transmitting over other
channels.}. At the beginning of each slot, a secondary user with
data to transmit chooses a set of channels to sense. A spectrum
sensor (\eg an energy detector) is used to detect the states of the
chosen channels. Based on the sensing outcomes, the secondary user
decides which sensed channels to access. Due to hardware and energy
constraints, we assume that a secondary user can sense and access at
most $L$ ($1\leq L\leq N$) channels in a slot. At the end of the
slot, the receiver acknowledges a successful transmission. The basic
slot structure is illustrated in Fig.~\ref{fig:slot}.

Our goal is to develop an optimal OSA strategy for the secondary
user, which sequentially determines which channels in the spectrum
to sense, how to design the spectrum sensor, and whether to access
based on the imperfect sensing outcomes. The design objective is to
maximize the throughput of the secondary user during a desired
period of $T$ slots under the constraint that the probability of
collision $P_n(t)$ perceived by the primary network in any channel
$n$ and slot $t$ is capped below a pre-determined threshold $\zeta$,
\ie
\begin{equation}\label{constraint}
P_n(t) \,\defeq\, \Pr\{\Phi_n(t) = 1\,|\,S_n(t) = 0\} \leq \zeta,
\quad \forall n, t,
\end{equation}
where $\Phi_n(t) \in \{0\mbox{ (no access), } 1\mbox{ (access)}\}$
denotes the access decision of the secondary user.

\noindent{\bf Remarks:}
\begin{enumerate}
  \item We assume that the transition probabilities of the
  SOS are known or have been learned. In Section
  \ref{sec:simulation}, we study the robustness of the
  optimal OSA design to a mismatched Markov model. For the case
  where the SOS dynamics are unknown, formulations and algorithms for
  POMDP with an unknown model exist in the literature \cite{Aberdeen:03TR} and can be applied to
  this problem, but is beyond the scope of this paper.

  \item We use the {\em conditional} probability of collision $P_n(t)$ in
  the design constraint and impose the collision constraint on any
  channel $n$ and slot $t$. This ensures that a primary user experiences collisions no more
  than $\zeta \times 100\%$ of its {\em transmission time} regardless of where and when it transmits.
  Note that if the unconditional probability of collision $\Pr\{\Phi_n(t) = 1, S_n(t) = 0\}$
  is adopted, the constraint depends on the traffic load of primary users in channels chosen
  by the secondary users; primary users who have light traffic load may not be as well protected as
  those with heavy traffic load.

  \item We assume that secondary users exploit spectrum
  opportunities independently and selfishly. That is, secondary
  users do not exchange their information on the SOS and
  everyone aims to maximize its own throughput without taking
  into consideration the interactions among secondary users. This assumption is
  suitable for secondary ad hoc networks where there is no
  central coordinator or dedicated control/communication channel. The
  secondary network can adopt a carrier sensing mechanism to
  avoid collisions among competing secondary users as detailed in \cite{Zhao&etal:05DySPAN,Zhao&etal:06JSAC}.
  We point out that such selfish decisions may not be optimal in terms of
  network-level throughput. Nevertheless, this formulation allows us
  to focus on the basic components of OSA and highlight the
  interaction among them.

\end{enumerate}


\section{Constrained POMDP Formulation}\label{sec:formulation}
Integrated in the optimal design of OSA are three basic components:
a spectrum sensor, a sensing strategy, and an access strategy. In
this section, we develop a decision-theoretic framework for the
optimal joint design based on the theory of POMDP. We focus first on
the single-channel sensing case where the secondary user can only
sense and access one channel in each slot ($L = 1$). Extensions to
multi-channel sensing scenarios are detailed in Section
\ref{sec:multiple}.

\subsection{Spectrum Sensor}\label{sec:sensor}
Suppose that channel $n$ is chosen in slot $t$. The spectrum sensor
detects the presence of primary users in this channel by performing
a binary hypothesis test:
\begin{equation}\label{H}
\begin{split}
&\Hc_0 : S_n(t) = 1 \mbox{~(idle)} \\
\mbox{~~~vs.~~~}&\Hc_1 : S_n(t) = 0 \mbox{~(busy)}.
\end{split}
\end{equation}
Let $\Theta_n(t) \in \{0\mbox{ (busy), } 1\mbox{ (idle)}\}$ denote
the sensing outcome (\ie the result of the binary hypothesis
test). 
The performance of the spectrum sensor is characterized by the PFA
$\epsilon_n(t)$ and the probability of miss detection (PM)
$\delta_n(t)$:
\begin{subequations}\label{ed}
\begin{align}
\epsilon_n(t) \,\defeq\, \Pr\{\mbox{decide }\Hc_1\,|\,\Hc_0\mbox{ is
true}\}= \Pr\{\Theta_n(t) = 0\,|\,S_n(t) = 1\},\\
\delta_n(t) \,\defeq\, \Pr\{\mbox{decide }\Hc_0\,|\,\Hc_1\mbox{ is
true}\} = \Pr\{\Theta_n(t) = 1\,|\,S_n(t) = 0\}.
\end{align}
\end{subequations}
For a given PFA $\epsilon_n(t)$, the largest achievable PD, denoted
as $P^{(n)}_{D,\max}(\epsilon_n(t))$, can be attained by the optimal
NP detector with the constraint that the PFA is no larger than
$\epsilon_n(t)$ or an optimal Bayesian detector with a suitable set
of risks \cite[Sec.~2.2.1]{Trees:01book}. All operating points
$(\epsilon, \delta)$ above the best ROC curve $P^{(n)}_{D,\max}$ are
thus infeasible.

\begin{figure}[htb]
\centerline{ \scalefig{0.7}\epsfbox{ 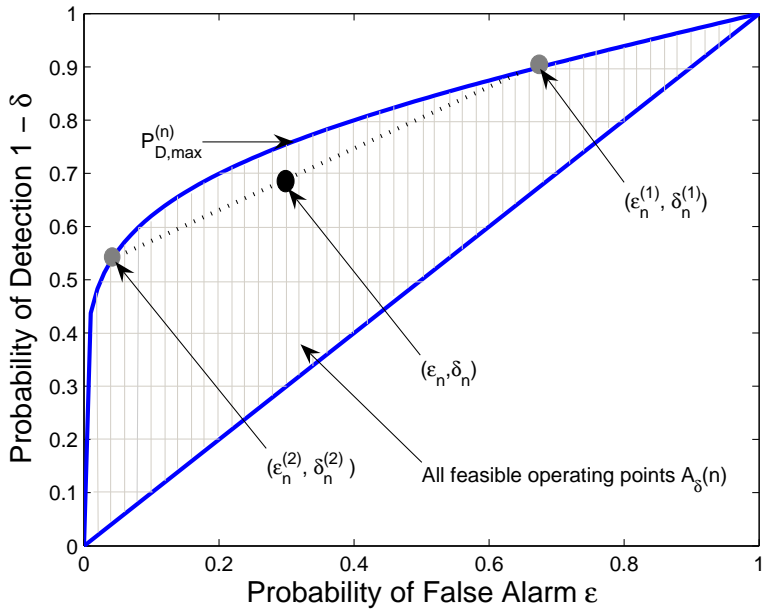}}
\caption{Illustration of the set $\mathbb{A}_\delta(n)$ of all
feasible sensor operating points $(\epsilon_n, \delta_n)$.
($\delta_n^{(i)} = 1 - P^{(n)}_{D,\max}(\epsilon_n^{(i)})$, $i =
1,2$) } \label{fig:ROCfeasible}
\end{figure}

Let $\mathbb{A}_\delta(n) \,\defeq\, \{(\epsilon, \delta): 0\leq
\epsilon \leq 1 - \delta \leq P^{(n)}_{D,\max}(\epsilon)\}$ denote
all feasible operating points of the spectrum sensor\footnote{Since
the two hypotheses in \eqref{H} play a symmetric role, we have
assumed, without loss of generality, that the PD is no smaller than
the PFA, \ie $1 - \delta \geq \epsilon$.}. As illustrated in
Fig.~\ref{fig:ROCfeasible}, the best ROC curve $P^{(n)}_{D,\max}$
achieved by the optimal NP detector forms the upper boundary of the
feasible set $\mathbb{A}_\delta(n)$. We also note that every sensor
operating point $(\epsilon_n,\delta_n)$ below the best ROC curve
lies on a line that connects two boundary points and hence can be
achieved by randomizing between two optimal NP detectors with
properly chosen constraints on the PFA
\cite[Sec.~2.2.2]{Trees:01book}. For example, the operating point
$(\epsilon_n,\delta_n)$ as shown in Fig.~\ref{fig:ROCfeasible} can
be achieved by applying the optimal NP detector under the constraint
of PFA $\leq \epsilon_n^{(1)}$ with probability $ p =
\frac{\epsilon_n - \epsilon_n^{(2)}}{\epsilon_n^{(1)} -
\epsilon_n^{(2)}}$ and the optimal NP detector under the constraint
of PFA $\leq\epsilon_n^{(2)}$ with probability $1-p$. Therefore, the
design of spectrum sensor is reduced to the choice of a desired
sensor operating point in $\mathbb{A}_\delta(n)$.

The design of the optimal NP detector is a well-studied classic
problem, which is not the focus of this paper. Our objective is to
define the criterion and the constraint under which the spectrum
sensor should be designed, equivalently, to find the optimal sensor
operating point $(\epsilon_n^*(t), \delta_n^*(t))
\in\mathbb{A}_\delta(n)$ to achieve the best tradeoff between false
alarm and miss detection. Note that the optimal sensor operating
point may vary with time (see Section \ref{sec:cmp} for an example.)

As discussed in Section \ref{sec:intro}, if the secondary user
completely trusts the sensing outcomes in making access decisions,
false alarms result in wasted spectrum opportunities while miss
detections lead to collisions with primary users. To optimize the
performance of the secondary user while limiting its interference to
the primary network, we need to carefully design the spectrum sensor
by considering its impact on the MAC layer performance in terms of
throughput and collision probability. Further, the spectrum access
decisions should be made by taking into account the sensor operating
characteristics. A joint design of the PHY layer spectrum sensor and
the MAC layer access strategy is thus necessary to achieve
optimality.



\subsection{Sensing and Access Strategies}\label{sec:sensing/access}
In each slot, a sensing strategy decides which channel in the
spectrum to sense, and an access strategy determines whether to
access given the sensing outcome\footnote{An alternative formulation
of the joint design is to combine the spectrum sensor with the
access strategy. In this case, the access decision is made directly
based on the channel measurements. It can be readily shown that this
formulation is equivalent to the one adopted here.}. Below we
illustrate the sequence of operations in each slot.

At the beginning of slot $t$, the SOS transits to $\Sbf(t) =
[S_1(t),\ldots,S_N(t)]$ according to the transition probabilities of
the underlying Markov process. The secondary user first chooses a
channel $a(t) \in \mathbb{A}_s \,\defeq\, \{1,\ldots,N\}$ to sense
and a feasible sensor operating point $(\epsilon_a(t),
\delta_a(t))\in \mathbb{A}_\delta(a(t))$. It then determines whether
to access $\Phi_a(t)\in \{0\mbox{ (no access), } 1\mbox{
(access)}\}$ by taking into account the sensing outcome
$\Theta_a(t)\in \{0\mbox{ (busy), } 1\mbox{ (idle)}\}$ provided by
the spectrum sensor that is designed according to the chosen
operating point $(\epsilon_a(t), \delta_a(t))$. A collision with
primary users happens when the secondary user accesses a busy
channel. At the end of this slot, the receiver acknowledges a
successful transmission $K_a(t) \in \{0 \mbox{ (no ACK), }1\mbox{
(ACK)}\}$. We assume that the ACK is error-free\footnote{Note that
the ACK is sent after the success reception of data. Hence, the
channel over which the ACK is transmitted is ensured to be idle in
this slot.}.

\subsection{Constrained POMDP Formulation}\label{sec:POMDP}
We show here that the joint design of OSA can be formulated as a
constrained POMDP with states, actions, transition probabilities,
observations, and reward structure defined as follows.

\noindent{\em State Space}~~~ The system state is given by the SOS
of the primary network. The state space is thus $\mathbb{S} =
\{0,1\}^N$.

\noindent{\em Action Space}~~~ In each slot $t$, the secondary user
needs to decide which channel to sense, which sensor operating point
to choose, and whether to access. Hence, the action in the POMDP
formulation consists of three parts: a sensing decision $a(t)\in
\mathbb{A}_s$, a spectrum sensor design $(\epsilon_a(t),
\delta_a(t))\in \mathbb{A}_\delta(a(t))$, and an access decision
$\Phi_a(t)\in\{0,1\}$.

\noindent{\em Transition Probabilities}~~~ The transition
probabilities of the SOS are given by $\{P_{\sbf,\sbf'}\}$, which
are determined by the primary traffic.

\noindent{\em Observation Space}~~~ As will become clear later,
optimal channel selection for opportunity tracking relies on the
exploitation of the statistical information on the SOS provided by
the observation history of the secondary users. To ensure
synchronous hopping in the spectrum without introducing extra
control message exchange, the secondary user and its desired
receiver must have the same history of observations so that they
make the same channel selection decisions. Since sensing errors may
cause different sensing outcomes at the transmitter and the
receiver, the acknowledgement $K_a(t) \in \{0,1\}$ should be used as
the common observation in each slot.

\noindent{\em Reward}~~~A nature definition of the reward is the
number of bits that can be delivered by the secondary user, which is
assumed to be proportional to the channel bandwidth. Given sensing
action $a(t)$ and access action $\Phi_a(t)$, the immediate reward
$R_{K_a(t)}$ can be defined as
\begin{equation}\label{R}
R_{K_a(t)} = K_a(t) B_a = S_a(t)\Phi_a(t)B_a.
\end{equation}
Hence, the expected total reward of the POMDP represents overall
throughput, the expected total number of bits that can be delivered
by the secondary user in $T$ slots.

\noindent{\em Belief Vector}~~~ Due to partial spectrum monitoring
and sensing errors, a secondary user cannot directly observe the
true SOS. It can, however, infer the SOS from its decision and
observation history. As shown in \cite{Smallwood&Sondik:73OR}, the
statistical information on the SOS provided by the entire decision
and observation history can be encapsulated in a belief vector
$\Lambdabf(t) \,\defeq\, \{\lambda_{\sbf}(t)\}_{\sbf \in
\mathbb{S}}\in\Pi(\mathbb{S})$, where $\lambda_{\sbf}(t)\in[0,1]$
denotes the conditional probability (given the decision and
observation history) that the SOS is $\sbf \in \mathbb{S}$ at the
beginning of slot $t$ {\em prior to} the state transition, and
\begin{equation}
\Pi(\mathbb{S}) \,\defeq\, \left\{ \{\lambda_{\sbf}\}_{\sbf \in
\mathbb{S}}: \lambda_\sbf \in[0,1], \sum_{\sbf\in\mathbb{S}}
\lambda_\sbf = 1\right\}
\end{equation}
denotes the belief space which includes all possible probability
mass functions (PMF) on the state space $\mathbb{S}$. Given belief
vector $\Lambdabf(t)$, the distribution of the system state $S(t)$
in slot $t$ after the state transition is then given by
\begin{equation}\label{a_priori}
\Pr\{\Sbf(t) = \sbf\} = \sum_{\sbf'\in\mathbb{S}}
\lambda_{\sbf'}(t)P_{\sbf',\sbf}, \qquad \forall \sbf\in\mathbb{S}.
\end{equation}

\noindent{\em Policy}~~~ A joint design of OSA is given by policies
of the above POMDP. Specifically, a sensing policy $\pi_s$ specifies
a sequence of functions, each mapping a belief vector $\Lambdabf(t)
\in\Pi(\mathbb{S})$ at the beginning of slot $t$ to a channel $a(t)
\in \mathbb{A}_s$ to be sensed in this slot: $\pi_s =
[\mu_s(1),\ldots,\mu_s(T)]$, where $\mu_s(t): \Pi(\mathbb{S})
\rightarrow \mathbb{A}_s$. Since the optimal policy for a
finite-horizon POMDP is generally non-stationary, functions
$\{\mu_s(t)\}_{t=1}^T$ are not identical. Similarly, a sensor
operating policy $\pi_\delta$ specifies, in each slot $t$, a
spectrum sensor design $(\epsilon_a(t), \delta_a(t))
\in\mathbb{A}_\delta(a(t))$ based on the current belief vector
$\Lambdabf(t)$ and the chosen channel $a(t)$. An access policy
$\pi_c$ specifies an access decision $\Phi_a(t) \in \{0,1\}$ in each
slot $t$ based on the current belief vector $\Lambdabf(t)$ and the
sensing outcome $\Theta_a(t) \in \{0,1\}$.

The above defined policies are deterministic. For unconstrained
POMDPs, there always exist deterministic optimal policies. For
constrained POMDPs, however, we may need to resort to randomized
policies to achieve optimality. A randomized sensing policy $\pi_s$
defines a sequence of functions, each mapping a belief vector
$\Lambdabf(t)$ to a PMF on the set $\mathbb{A}_s$ of channels, and a
randomized sensor operating policy $\pi_\delta$ defines the mapping
from $\Lambdabf(t)$ to a probability density function (PDF) on the
set $\mathbb{A}_\delta(a(t))$ of feasible sensor operating points. A
randomized access policy $\pi_c$ maps $\Lambdabf(t)$ and sensing
outcome $\Theta_a(t)$ to a transmission probability in each slot
$t$. In other words, the actions chosen in a randomized policy are
probability distributions. Due to the uncountable space of
probability distributions, randomized policies are usually
computationally prohibitive.

\noindent{\em Objective and Constraint}~~~ We aim to develop the
optimal joint design of OSA $\{\pi_\delta^*,\pi_s^*,\pi_c^*\}$ that
maximizes the expected total number of bits that can be delivered by
the secondary user (\ie the expected total reward of the POMDP) in
$T$ slots under the collision constraint given in
\eqref{constraint}:
\begin{equation}\label{obj}
\begin{split}
&\{\pi_\delta^*,\pi_s^*,\pi_c^*\} =
\arg\max_{\pi_\delta,\pi_s,\pi_c}
\mathbb{E}_{\{\pi_\delta,\pi_s,\pi_c\}}\left[\left.\sum_{t=1}^T
R_{K_a(t)}\right| \Lambdabf(1) \right] \\
&\mbox{s.t. } P_a(t) =  \Pr\{\Phi_a(t) = 1\,|\,S_a(t) = 0\}\leq
\zeta, \qquad \forall a, t,
\end{split}
\end{equation}
where $\mathbb{E}_{\{\pi_\delta,\pi_s,\pi_c\}}$ represents the
expectation given that policies $\{\pi_s,\pi_\delta,\pi_c\}$ are
employed, $P_a(t)$ is the probability of collision perceived by the
primary network in chosen channel $a(t)$ and slot $t$, and
$\Lambdabf(1)$ is the initial belief vector, which can be set to the
stationary distribution of the underlying Markov process if no
information on the initial SOS is available.

We consider in \eqref{obj} the non-trivial case where the
conditional collision probability $P_a(t)$ is well-defined, \ie
$\Pr\{S_a(t) = 0\} > 0$. Note that $\Pr\{S_a(t) = 0\} = 0$ (or 1)
implies that the system state $S_a(t)$ is known  based on the
current belief vector $\Lambdabf(t)$. In this case, the optimal
access decision is straightforward, and the design of the spectrum
sensor becomes unnecessary since the channel state is already known.

\section{Separation Principle for Optimal OSA}\label{sec:one}
In this section, we solve the constrained POMDP given in \eqref{obj}
to obtain the optimal joint design of OSA. Specifically, we
establish a separation principle that reveals the optimality of
deterministic policies and leads to closed-form optimal design of
the spectrum sensor and the access strategy. It also allows us to
characterize quantitatively the interaction between the PHY layer
sensor operating characteristics and the MAC layer access strategy.

\subsection{Optimality Equation}\label{sec:opt_eq}
The first step to solving \eqref{obj} is to express the objective
and the constraint explicitly as functions of the actions. We
establish first the optimality of deterministic sensing and sensor
operating policies, which significantly simplifies the action space.

\vspace{0.5em}

\noindent{\em Optimality of deterministic policies}~~~ In
Proposition \ref{prop:deterministic}, we show that it is sufficient
to consider deterministic sensing and sensor operating policies in
the optimal joint design of OSA.

\begin{proposition}\label{prop:deterministic}
For the optimal joint design of OSA given by \eqref{obj}, there
exist deterministic optimal sensing and sensor operating policies.
\end{proposition}

\begin{proof}
The proof is based on the concavity of the best ROC curve and the
fact that the collision constraint is imposed on every channel. See
details in Appendix A.
\end{proof}

As a result of Proposition \ref{prop:deterministic}, the secondary
user needs to choose, in each slot\footnote{Time index $t$ will be
omitted for notation convenience.}, a channel $a\in\mathbb{A}_s$ to
sense, a feasible sensor operating point $(\epsilon_a, \delta_a) \in
\mathbb{A}_\delta(a)$, and a pair of transmission probabilities
$(f_a(0),f_a(1))$, where
\begin{equation*}
f_a(\theta) \,\defeq\, \Pr\{\Phi_a = 1\,|\, \Theta_a = \theta \}\in
[0,1]
\end{equation*}
is the probability of accessing channel $a$ given sensing outcome
$\Theta_a = \theta\in\{0,1\}$. The composite action space is then
given by
\begin{equation}
\mathbb{A}\,\defeq\, \{(a,(\epsilon_a, \delta_a), (f_a(0), f_a(1))):
a\in\mathbb{A}_s, (\epsilon_a, \delta_a) \in \mathbb{A}_\delta(a),
(f_a(0), f_a(1)) \in [0,1]^2\}.
\end{equation}

\vspace{0.5em}

\noindent{\em Objective function}~~~ Let $V_t(\Lambdabf(t))$ be the
value function, which represents the maximum expected reward that
can be obtained starting from slot $t$ ($1\leq t\leq T$) given
belief vector $\Lambdabf(t)$ at the beginning of slot $t$. Given
that the secondary user takes action $A = \{a,(\epsilon_a,
\delta_a), (f_a(0), f_a(1))\} \in\mathbb{A}$ and observes
acknowledgement $K_a = k$, the reward that can be accumulated
starting from slot $t$ consists of two parts: the immediate reward
$R_{K_a}= kB_a$ and the maximum expected future reward
$V_{t+1}(\Lambdabf(t+1))$, where
\begin{equation*}
\Lambdabf(t+1) \,\defeq\, \{\lambda_\sbf(t+1)\}_{\sbf\in\mathbb{S}}
=\Tc(\Lambdabf(t) \,|\,A, k)
\end{equation*}
represents the updated knowledge of the SOS after incorporating the
action $A$ and the acknowledgement $k$ in slot $t$. Averaging over
all possible states $\sbf\in\mathbb{S}$ and acknowledgements
$k\in\{0,1\}$ and maximizing over all actions $A\in\mathbb{A}$, we
arrive at the following optimality equation
\begin{subequations}\label{V1}
\begin{align}
V_t(\Lambdabf(t)) &= \max_{A\in\mathbb{A}} \sum_{\sbf\in\mathbb{S}}
\sum_{\sbf' \in\mathbb{S}} \lambda_{\sbf'}(t)P_{\sbf',\sbf}
\sum_{k=0}^1 U_{\sbf,k}(A) \left[ kB_a  +
V_{t+1}(\Tc(\Lambdabf(t)\,|\,A, k))\right], ~~1\leq t < T, \\
V_T(\Lambdabf(T)) &= \max_{A \in\mathbb{A}} \sum_{\sbf\in\mathbb{S}}
\sum_{\sbf' \in\mathbb{S}} \lambda_{\sbf'}(t)P_{\sbf',\sbf}
U_{\sbf,1}(A) B_a ,\label{VT1}
\end{align}
\end{subequations}
where $\sum_{\sbf' \in\mathbb{S}} \lambda_{\sbf'}(t)P_{\sbf',\sbf}$
is the distribution  of the SOS in slot $t$ (see \eqref{a_priori}),
and $U_{\sbf,k}(A)\,\defeq\, \Pr\{K_a = k\,|\,\Sbf = \sbf\}$ is the
conditional distribution of the acknowledgement given current state
$\sbf$ and action $A$. Since $K_a = S_a \Phi_a$, the conditional
distribution $U_{\sbf,k}(A)$ of the acknowledgement can be
calculated as
\begin{subequations}\label{U1}
\begin{align}
U_{\sbf,1}(A) & \,\defeq\, \Pr\{K_a = 1\,|\,\Sbf = \sbf\}
= \Pr\{S_a = 1\,|\,\Sbf = \sbf\}\Pr\{\Phi_a = 1\,|\,\Sbf = \sbf, S_a = 1\} \notag \\
&=1_{[s_a = 1]}\sum_{\theta=0}^1 \Pr\{\Theta_a = \theta\,|\,\Sbf =
\sbf\} f_a(\theta) = s_a[\epsilon_af_a(0) + (1-\epsilon_a)f_a(1)],\\
U_{\sbf,0}(A) &= 1 - U_{\sbf,1}(A),
\end{align}
\end{subequations}
where $1_{[x]}$ is the indicator function and $\Pr\{S_a = 1\,|\,\Sbf
= \sbf\} = 1_{[s_a = 1]}$ is given by the occupancy state $s_a$ of
channel $a$. Applying Bayes' rule, we obtain the updated belief
vector $\Lambdabf(t+1) =\Tc(\Lambdabf(t) \,|\,A, k) $ as
\begin{equation}\label{T}
\begin{split}
    \lambda_{\sbf}(t+1) = \frac{\sum_{\sbf'
\in\mathbb{S}} \lambda_{\sbf'}(t)P_{\sbf',\sbf} U_{\sbf,k}(A)}
{\sum_{\sbf \in\mathbb{S}}\sum_{\sbf' \in\mathbb{S}}
\lambda_{\sbf'}(t)P_{\sbf',\sbf} U_{\sbf,k}(A)}, \qquad
\sbf\in\mathbb{S}.
\end{split}
\end{equation}
We see from \eqref{T} that by adopting the acknowledgement $K_a$ as
their observation, the transmitter and the receiver will have the
same updated belief vector $\Lambdabf(t+1)$, which ensures that they
tune to the same channel in the next slot.

Note from \eqref{V1} that the action $A = \{a,(\epsilon_a,
\delta_a),(f_a(0),f_a(1))\}$ taken by the secondary user affects the
expected total reward in two ways: it acquires an immediate reward
$R_{K_a} = kB_a$ and transforms the current belief vector
$\Lambdabf(t)$ to a new one $\Lambdabf(t+1) = \Tc(\Lambdabf(t)
\,|\,A, k)$ which determines the future reward
$V_{t+1}(\Tc(\Lambdabf(t) \,|\,A, k))$. Hence, the function of the
secondary user's action is twofold: to exploit immediate spectrum
opportunities and to gain information on the SOS (characterized by
belief vector $\Lambdabf(t+1)$) so that more rewarding decisions can
be made in the future. As a consequence, the optimal joint design of
OSA should achieve the tradeoff between these two often conflicting
objectives. Myopic policies that aim solely at maximizing the
instantaneous throughput (\ie the expected immediate reward) without
considering future consequences are generally suboptimal.

\vspace{0.5em}

\noindent{\em Collision Constraint}~~~ The collision probability
$P_a(t)$ is determined by the sensor operating point
$(\epsilon_a,\delta_a)$ and the transmission probabilities $(f_a(0),
f_a(1))$:
\begin{align}\label{constraint1}
P_a(t) &\,\defeq\, \Pr\{\Phi_a(t) = 1\,|\,S_a(t) = 0\}\notag\\
& = \sum_{\theta = 0}^1 \Pr\{\Theta_a = \theta\,|\,S_a = 0\}
\Pr\{\Phi_a = 1\,|\, \Theta_a = \theta, S_a = 0 \} \notag\\
&= (1-\delta_a)f_a(0) + \delta_a f_a(1) \leq \zeta.
\end{align}

\vspace{0.5em}

In principle, by solving \eqref{V1} recursively (starting from the
last slot $T$ using \eqref{VT1}) under the constraint of
\eqref{constraint1}, we can obtain the maximum overall throughput
$V_1(\Lambdabf(1))$ of the secondary user and the corresponding
policies $\{\pi_s^*,\pi_\delta^*,\pi_c^*\}$. However, \eqref{V1} is
generally intractable due to the uncountable action space
$\mathbb{A}$.

\subsection{The Separation Principle}\label{sec:sepa1}

\begin{theorem} {\bf The Separation Principle for OSA with Single-Channel Sensing}\label{th:sep}

{\em The joint design of OSA given in \eqref{V1} can be carried out
in two steps without losing optimality.
\begin{itemize}
  \item Step 1: Choose the sensor operating policy $\pi_\delta$ and the
  access policy $\pi_c$ to maximize the instantaneous throughput subject
  to the collision constraint. Specifically, for any chosen channel $a$, the optimal sensor operating
  point $(\epsilon_a^*, \delta_a^*)$ and transmission probabilities $(f_a^*(0), f_a^*(1))$
  are given by
\begin{subequations}\label{s1}
\begin{align}
&\{(\epsilon_a^*, \delta_a^*), (f_a^*(0), f_a^*(1))\}
=\arg\max_{\substack{(\epsilon_a, \delta_a)\in\mathbb{A}_\delta(a)\\
(f_a(0),f_a(1)) \in [0,1]^2}} \mathbb{E}\left[\left.R_{K_a(t)}\,\right|\,\Lambdabf(t)\right] \notag\\
&\qquad\qquad\qquad\qquad\quad\quad
= \arg\max_{\substack{(\epsilon_a, \delta_a)\in\mathbb{A}_\delta(a)\\
(f_a(0),f_a(1)) \in [0,1]^2}} \epsilon_af_a(0) + (1-\epsilon_a)
f_a(1) \label{s1_obj}\\
&\mbox{s.t.}\quad P_a(t) = (1-\delta_a)f_a(0) + \delta_a f_a(1) \leq
\zeta.
\end{align}
\end{subequations}

  \item Step 2: Using the optimal sensor operating and access policies
    $\{\pi_\delta^*,\pi_c^*\}$ given by \eqref{s1},
    choose sensing policy to maximize the overall throughput.
  Specifically, the optimal sensing policy $\pi_s^*$ is given by
    \begin{equation}\label{s2}
        \pi_s^* = \arg\max_{\pi_s} \mathbb{E}_{\pi_s}
        \left[\left.\sum_{t=1}^T R_{K_a(t)}\right| \Lambdabf(1)
        \right].
    \end{equation}
\end{itemize}}
\end{theorem}

\vspace{0.5em}
\begin{proof}
The proof is based on the convexity of the value function
$V_t(\Lambdabf(t))$ with respect to the belief vector $\Lambdabf(t)$
and the structure of the conditional observation distributions
$U_{\sbf,k}(A)$. See Appendix B for details.
\end{proof}

The separation principle simplifies the optimal joint design of OSA
in two ways. First, it reveals that myopic policies, rarely optimal
for a general POMDP, are optimal for the design of the spectrum
sensor and the access strategy. We can thus obtain the optimal
spectrum sensor $(\epsilon_a^*, \delta_a^*) \in
\mathbb{A}_\delta(a)$ and the optimal transmission probabilities
$(f_a^*(0), f_a^*(1))\in [0,1]^2$ by solving a static optimization
problem given in \eqref{s1}. This allows us to characterize
quantitatively the interaction between the spectrum sensor and the
access strategy as given Proposition \ref{prop:f} and to obtain the
optimal joint design in closed-form as given in Theorem
\ref{th:opt}. While the proof is lengthy, there is an intuitive
explanation for this apparently surprising. We note that upon
receiving the ACK $K_a=1$, the secondary user knows exactly that the
chosen channel is idle. However, when $K_a = 0$ (no packet is
received), the secondary receiver cannot tell whether the chosen
channel is busy or not accessed. Hence, $K_a = 1$ provides the
secondary user with more information on the current SOS. We also
note that accessing the chosen channel maximizes not only the
instantaneous throughput but also the chance of receiving more
informative observation $K_a = 1$. Hence, getting immediate reward
and gaining information for more rewarding future decisions are no
longer conflicting here.

Second, the separation principle decouples the design of the sensing
strategy from that of the spectrum sensor and the access strategy.
Furthermore, it reduces the design of the sensing strategy from a
constrained POMDP \eqref{obj} to an unconstrained one with finite
action space \eqref{s2}. This is because the sensor operating points
and the transmission probabilities determined by \eqref{s1} have
ensured the collision constraint regardless of channel selections.
The optimal sensing policy is thus obtained by maximizing the
overall throughput without any constraint. Unconstrained POMDPs have
been well-studied. The optimal sensing policy can thus be readily
obtained by using computationally efficient solution procedures in
\cite{Sondik:71PhD,Smallwood&Sondik:73OR,Monahan:82MS,Cheng:88PhD}.

\subsection{Interaction between the PHY and the MAC Layers}\label{sec:pic}
Before solving for the optimal sensor operating and access policies,
we study the interaction between the PHY layer spectrum sensor and
the MAC layer access strategy.

We note that when the spectrum sensor at the PHY layer is given, the
separation principle still holds for the design of the sensing and
access strategies. The optimal access strategy for a given spectrum
sensor can thus be obtained.

\begin{proposition}\label{prop:f}
{\em Given a chosen channel $a$ and a feasible sensor operating
point $(\epsilon_a, \delta_a)$, the optimal transmission
probabilities $(f_a^*(0), f_a^*(1))$ are given by
\begin{equation}\label{f}
    (f_a^*(0), f_a^*(1)) =
    \begin{cases} (\frac{\zeta - \delta_a}{ 1 -
    \delta_a},1),\quad & \delta_a < \zeta, \\
    (0,1), & \delta_a = \zeta,\\
    (0,\frac{\zeta}{\delta_a}), & \delta_a > \zeta.
    \end{cases}
\end{equation}
}
\end{proposition}

\vspace{0.5em}

\begin{proof}
The proof is based on the separation principle \eqref{s1} and the
fact that all feasible operating points lie above the line $1 -
\delta_a = \epsilon_a$. See details in Appendix C.
\end{proof}

\begin{figure}[htb]
\centerline{
\begin{psfrags}
\psfrag{B}[l]{\footnotesize $\delta_a > \zeta$: conservative}
\psfrag{A}[l]{\footnotesize $\delta_a < \zeta$: aggressive}
\scalefig{0.7}\epsfbox{ 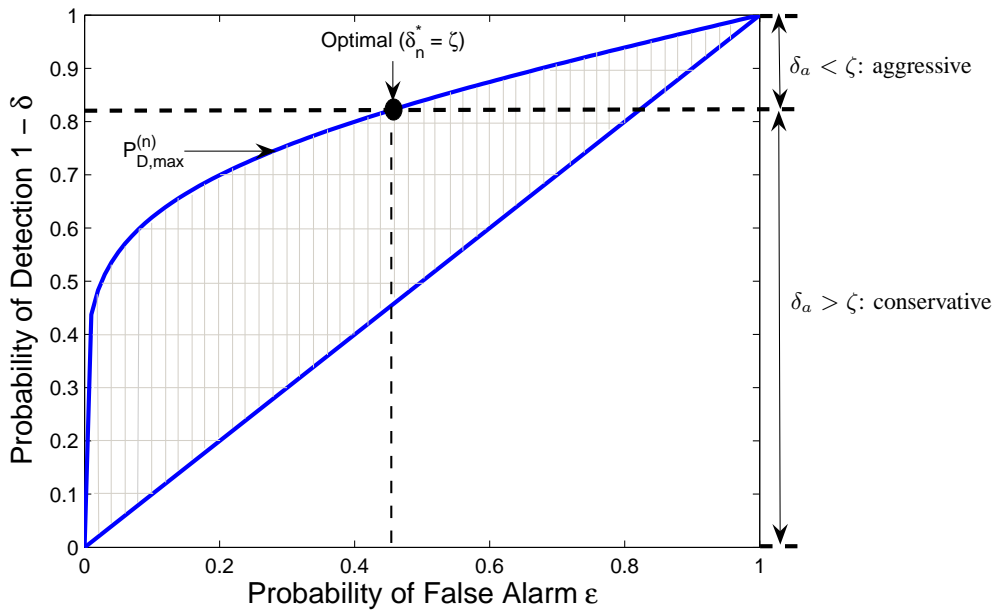}
\end{psfrags}}
\caption{Illustration of conservative and aggressive
regions.}\label{fig:ROCopt}
\end{figure}

As seen from Proposition \ref{prop:f}, randomized access policies
are necessary to achieve optimality when $\delta_a \neq \zeta$.
Moreover, Proposition \ref{prop:f} quantitatively characterizes the
impact of the sensor performance $\delta_a$ on the optimal access
strategy $(f_a^*(0), f_a^*(1))$. As illustrated in
Fig.~\ref{fig:ROCopt}, the set $\mathbb{A}_\delta(a)$ of feasible
sensor operating points can be partitioned into two regions: the
``conservative'' region ($\delta_a > \zeta$) and the ``aggressive''
region ($\delta_a < \zeta$). When $\delta_a > \zeta$, with high
probability, the spectrum sensor detects a busy channel as idle (\ie
a miss detection occurs). Hence, the access policy should be
conservative to ensure that the collision probability is capped
below $\zeta$. Specifically, even when the sensing outcome $\Theta_a
= 1$ indicates an idle channel, the secondary user should only
transmit with probability $\frac{\zeta}{\delta_a}<1$. When the
channel is sensed as busy $\Theta_a = 0$, the user should always
refrain from transmission. On the other hand, when $\delta_a <
\zeta$, the probability of false alarm is high; the spectrum sensor
is likely to overlook an opportunity. Hence, the secondary user
should adopt an aggressive access policy: always transmit when the
channel is sensed as idle and transmit with probability $\frac{\zeta
- \delta_a}{1-\delta_a} > 0$ even when the sensing outcome indicates
a busy channel. When $\delta_a = \zeta$, the access policy is to
simply trust the sensing outcome: $\Phi_a = \Theta_a$. We will show
in Section \ref{sec:picd} that the splitting point $\delta_a =
\zeta$ on the best ROC curve $P_{D,\max}^{(a)}$ is the optimal
sensor operating point.

Similar to Proposition \ref{prop:f}, we can quantitatively study the
impact of the access strategy on the spectrum sensor design by
solving \eqref{s1} for the optimal sensor operating points when the
transmission probabilities are given. This result is omitted to
avoid unnecessary repetition. Details can be found in
\cite{Chen&etal:07TR-02}.

%
%
%
%
%
%

\subsection{Optimal Joint Design of Spectrum Sensor and Access
Policy}\label{sec:picd}

Optimizing \eqref{f} over all feasible sensor operating points, we
obtain an explicit optimal design for the spectrum sensor and a
closed-form deterministic optimal access policy in Theorem
\ref{th:opt}.

\begin{theorem}\label{th:opt}
{\em  For any chosen channel $a$ in any slot, the optimal sensor
should adopt the optimal NP detector with constraint $\delta_a^* =
\zeta$ on the PM. Correspondingly, the optimal access policy is to
trust the sensing outcome given by the spectrum sensor, \ie
$f^*_a(0) = 0$ and $f^*_a(1) = 1$. }
\end{theorem}

\begin{proof}
The proof of Theorem \ref{th:opt} exploits the convexity of the set
$\mathbb{A}_\delta$ of feasible sensor operating points, which
follows directly from the concavity of the best ROC curve
\cite{Trees:01book}. See Appendix D for details.
\end{proof}

We find that the optimal sensor operating point coincides with the
splitting point $\delta_a^* = \zeta$ of the ``conservative'' region
and the ``aggressive'' region on the best ROC curve (see
Fig.~\ref{fig:ROCopt}). This indicates that at $\delta^*_a = \zeta$,
the best tradeoff between false alarm and miss detection is achieved
and the access policy does not need to be conservative or
aggressive. We thus have a simple and deterministic optimal access
policy: trust the sensing outcome $\Phi_a = \Theta_a$, \ie access if
and only if the channel is sensed to be available. Summarized below
are the properties of the optimal sensor operating and access
policies given in Theorem \ref{th:opt}.

\noindent{\em Properties 1:~~~The optimal spectrum sensor design and
the optimal access policy  are}
\begin{itemize}
  \item[P1.1]  {\em time-invariant and belief-independent}.
  \item[P1.2]  {\em model-independent}.
\end{itemize}

As a result of P1.1, the spectrum sensor can be configured off-line,
and there is no need to calculate and store the optimal transmission
probabilities, leading to significant reduction in both
implementation complexity and memory requirement. The second
property is that the optimal design of the spectrum sensor and the
access strategy  does not require the knowledge of the transition
probabilities of the underlying Markov process. Since the
probability of collision \eqref{constraint1} is solely determined by
the sensor operating and access policies, P1.2 indicates that the
collision constraint on the joint OSA design can be ensured
regardless of the accuracy of the Markovian model used by the
secondary user. In other words, the primary network is not affected
by the inaccurate model adopted by the secondary user. Model
mismatch only affects the performance of the secondary user (see
Fig.~\ref{fig:mc_error} for a simulation example).

\subsection{Optimal Sensing Policy}\label{sec:pis}
As revealed by the separation principle, the optimal sensing policy
can be obtained by solving an unconstrained POMDP with finite action
space $\mathbb{A}_s$. Specifically, by applying the optimal spectrum
sensor design and the optimal access policy given in Theorem
\ref{th:opt} to \eqref{V1}, we simplify the optimality equation as
\begin{subequations}\label{V1b}
\begin{align}
V_t(\Lambdabf(t)) &= \max_{a\in\mathbb{A}_s}
\sum_{\sbf\in\mathbb{S}} \sum_{\sbf' \in\mathbb{S}}
\lambda_{\sbf'}(t)P_{\sbf',\sbf}\sum_{k=0}^1U_{\sbf,k}(a)
[kB_a + V_{t+1}(\Tc(\Lambdabf(t)\,|\,a,k))], \qquad 1\leq t < T,\\
V_T(\Lambdabf(T)) &= \max_{a\in\mathbb{A}_s}
\sum_{\sbf\in\mathbb{S}} \sum_{\sbf' \in\mathbb{S}}
\Lambda_{\sbf'}(t)P_{\sbf',\sbf} U_{\sbf,k}(a) B_a.
\end{align}
\end{subequations}
By applying $f^*_a(0) = 0$ and $f^*_a(1) = 1$ to \eqref{U1}, we
obtain the conditional observation probability $U_{\sbf,1}(a)$ as
\begin{equation}
U_{\sbf,1}(a) = s_a(1-\epsilon_a^*),\quad U_{\sbf,0}(a) = 1 -
U_{\sbf,1}(a),
\end{equation}
where $\epsilon_a^*$ is the PFA associated with the PD $1 - \delta^*
= 1 - \zeta$ on the best ROC curve $P_{D,\max}^{(a)}$. The updated
belief vector $\Tc(\Lambdabf(t)\,|\,a,k)$ can be obtained by
substituting $U_{\sbf,k}(A)$ in \eqref{T} with $U_{\sbf,k}(a)$.

It is shown in \cite{Smallwood&Sondik:73OR} that the value function
of an unconstrained POMDP with finite action space is piece-wise
linear and can be solved via linear programming. We can thus use the
existing computationally efficient algorithms \cite{Sondik:71PhD,
Monahan:82MS, Cheng:88PhD} to solve \eqref{V1} for the optimal
sensing policy.

Although myopic sensor operating and access policies are shown to be
optimal for the joint design of OSA (see the separation principle),
myopic sensing policy is suboptimal in general. Interestingly, it
has been shown in \cite{Zhao&Krishnamachari:07CogNet} that the
myopic sensing policy is optimal when the SOS evolves independently
and identically across channels. When the channel occupancy states
are correlated, the myopic approach can serve as a suboptimal
solution with reduced complexity.

\subsection{Simulation Examples}\label{sec:simulation}
\begin{figure}[htb]
\centerline{
\begin{psfrags}
\psfrag{A}[c]{ $0$} \psfrag{B}[c]{ $1$} \psfrag{A1}[c]{(busy)}
\psfrag{B1}[c]{(idle)} \psfrag{a}[c]{ $\alpha_n$} \psfrag{b}[l]{
$\beta_n$} \psfrag{a1}[r]{ $1-\alpha_n$} \psfrag{b1}[c]{
$1-\beta_n$} \scalefig{0.4}\epsfbox{ 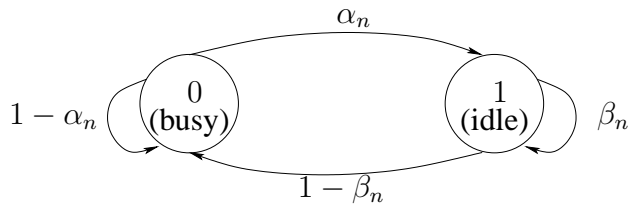}
\end{psfrags}}
\caption{The Markov channel model.} \label{fig:MC}
\end{figure}

Here we provide simulation examples to study different factors that
affect the optimal joint design of OSA. We consider $N = 3$
channels, each with bandwidth $B_n = 1$. While the separation
principle applies to arbitrarily correlated SOS, we consider here
the case where the SOS evolves independently but not identically
across these three channels for simplicity. As illustrated in
Fig.~\ref{fig:MC}, the SOS dynamics are given by the transition
probabilities $\alphabf \,\defeq\, [\alpha_1, \alpha_2, \alpha_3]$
and $\betabf\,\defeq\, [\beta_1,\beta_2,\beta_3]$, where $\alpha_n$
denotes the probability that channel $n$ transits from state $0$
(busy) to state $1$ (idle), and $\beta_n$ denotes the probability
that channel $n$ stays in state $1$. In all figures, the transition
probabilities are given by $\alphabf = [0.2,0.4,0.6]$ and $\betabf =
[0.8,0.6,0.4]$. We assume that they remain unchanged in $T=10$
slots. The maximum allowable probability of collision is $\zeta =
0.05$. We use the normalized overall throughput
$V_1(\Lambdabf(1))/T$, where $\Lambdabf(1)$ is the stationary
distribution of the SOS, to evaluate the performance of the optimal
OSA design.

To illustrate the interaction between the PHY layer spectrum sensor
and the MAC layer access policy, we consider a simple spectrum
sensing scenario where the background noise and the primary signal
are modeled as white Gaussian processes. Let $\sigma_{n,0}^2$ and
$\sigma_{n,1}^2$ denote, respectively, the noise and the primary
signal power in channel $n$. At the beginning of each slot, the
spectrum sensor takes $M$ independent measurements $\Ybf_n
\,\defeq\, [Y_{n,1},\ldots,Y_{n,M}]$ from chosen channel $n$ and
performs the following binary hypothesis test:
\begin{equation}\label{Htest1}
\begin{split}
\Hc_0 (S_n = 1): & \qquad  \Ybf_n\sim\Nc(\mathbf{0}_M,\sigma^2_{n,0}\Ibf_M), \\
\mbox{vs.~~~}\Hc_1 (S_n = 0): &  \qquad
\Ybf_n\sim\Nc(\mathbf{0}_M,(\sigma^2_{n,1} + \sigma^2_{n,0})\Ibf_M),
\end{split}
\end{equation}
where $\Nc(\mathbf{0}_M,\sigma^2\Ibf_M)$ denotes the $M$-dimensional
Gaussian distribution with identical mean 0 and variance $\sigma^2$
in each dimension. An energy detector is optimal under the NP
criterion \cite[Sec.~2.6.2]{Trees:01book}:
\begin{equation}\label{test1}
||\Ybf_n||_2 = \sum_{i=1}^M Y_{n,i}^2 \gtrless^{\Hc_1}_{\Hc_0}
\eta_n.
\end{equation}
The PFA and the PM of the energy detector are given by
\cite[Sec.~2.6.2]{Trees:01book}:
\begin{align}\label{delta_epsilon}
\delta_n = \gamma\left(\frac{M}{2},
\frac{\eta_n}{2(\sigma_{n,0}^2+\sigma_{n,1}^2)}\right),
~~~~\epsilon_n = 1 - \gamma\left(\frac{M}{2},\frac{\eta_n}
{2\sigma_{n,0}^2}\right),
\end{align}
where $\gamma(m,a) = \frac{1}{\Gamma(m)}\int_0^a t^{m-1}e^{-t}\,dt$
is the incomplete gamma function. The optimal decision threshold
$\eta_n^*$ of the energy detector is chosen so that $\delta_n^* =
\zeta$. Unless otherwise mentioned, we assume that $M= 10$,
$\sigma_{n,0}^2 = \sigma_0^2 = 0$ dB, and $\sigma_{n,1}^2 =
\sigma_1^2 = 5$ dB for all channels $n = 1,\ldots,N$.

\subsubsection{Impact of Sensor Operating Characteristics}


\begin{figure}[h]
\centerline{
\begin{psfrags}
\psfrag{c}[l]{{\bf $\delta\uparrow$
$\epsilon\downarrow$}(conservative)} %
\psfrag{b}[r]{{\bf $\delta\downarrow$
$\epsilon\uparrow$}(aggressive)} \scalefig{0.7}\epsfbox{
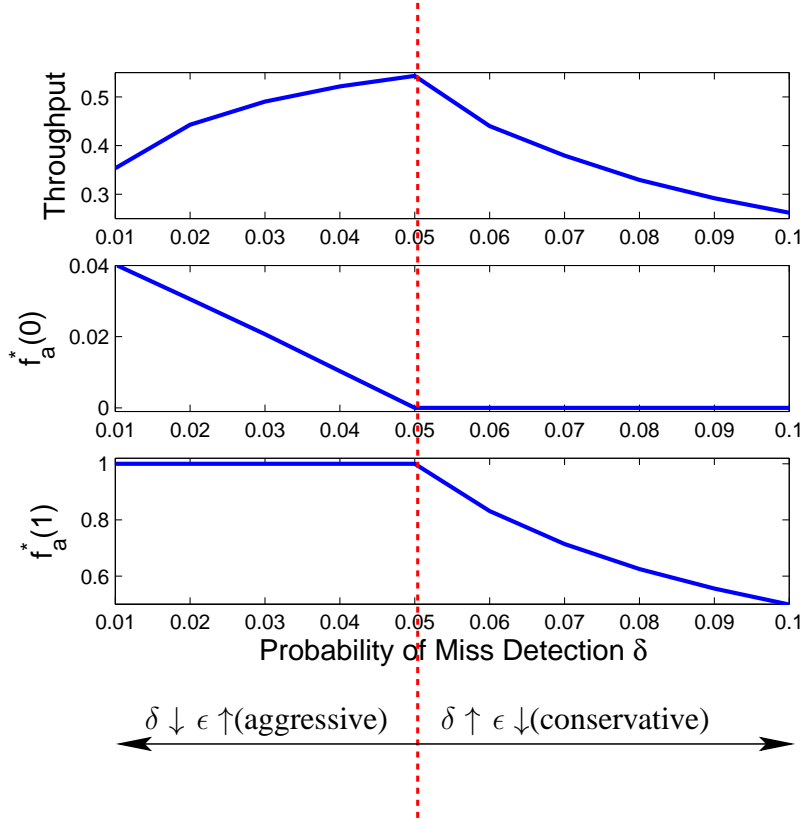}
\end{psfrags}}
\caption {The impact of sensor operating characteristics on the
performance of the optimal OSA design. } \label{fig:delta}
\end{figure}

Fig. \ref{fig:delta} shows the impact of sensor operating
characteristics on the secondary user's throughput and the optimal
access policy. The upper figure plots the maximum throughput
$V_1(\Lambdabf(1))/T$ vs. the PM $\delta$. The optimal transmission
probabilities $(f_a^*(0), f_a^*(1))$ are shown in the middle and the
lower figures, respectively. We can see that the maximum throughput
is achieved at $\delta^* = \zeta = 0.05$ and the transmission
probabilities change with $\delta$ as given by Theorem \ref{prop:f}.
Interestingly, the throughput curve is concave with respect to
$\delta$ in the ``aggressive'' region ($\delta < \zeta$) and convex
in the ``conservative'' region ($\delta > \zeta$). The performance
thus decays at a faster rate when the sensor operating point drifts
toward the ``conservative'' region. This suggests that miss
detections are more harmful to the OSA design than false alarms.

\subsubsection{Impact of the Number of Channel Measurements}

\begin{figure}[!htb]
\centerline{ \scalefig{0.7}\epsfbox{ 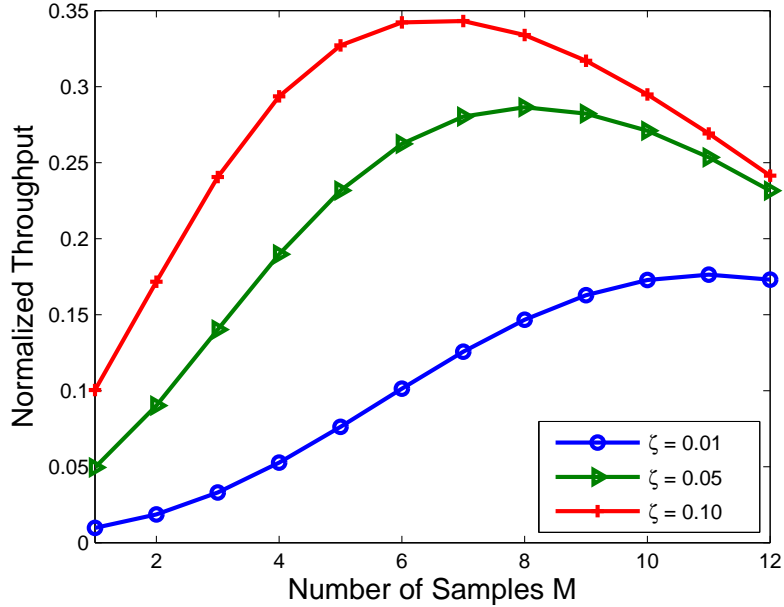}} \caption {The
impact of the number of channel measurements on the performance of
the optimal OSA design.} \label{fig:samples}
\end{figure}

In this example, we study the tradeoff between the spectrum sensing
time, which determines on the number $M$ of channel measurements
taken by the spectrum sensor, and the transmission time. Taking more
channel measurements can improve the fidelity of the sensing outcome
but will reduce the data transmission time and hence the number of
transmitted bits. We are thus motivated to study the throughput of
the secondary user as a function of $M$ for different maximum
allowable probabilities of collision $\zeta$. We assume that each
channel measurement takes $c = 5\%$ of a slot time. The transmission
time is thus given by $1-Mc = 1 - 0.05M$. Assuming that the number
of bits that can be transmitted by the secondary user is
proportional to both the channel bandwidth and the transmission
time, we modify the immediate reward \eqref{R} of the POMDP to
$R_{K_a} = (1-Mc)K_aB_a$.

Fig.~\ref{fig:samples} shows that the throughput of the secondary
user increases and then decreases with the number $M$ of channel
measurements. Note that the PM is a function of the number $M$ of
channel measurements and the detection threshold $\eta_a^*$ of the
energy detector (as seen from \eqref{delta_epsilon}). When the PM is
fixed to be $\delta_a^* = \zeta$ according to the separation
principle, the detection threshold $\eta_a^*$ increases with $M$,
and hence the PFA $\epsilon_a^*$ decreases with $M$. As a
consequence, when $M$ is small, the throughput of the secondary user
is limited by the large PFA. On the other hand, when $M$ is large,
the PFA is reduced at the expense of less transmission time in each
slot, which also leads to low throughput. We also observe that the
optimal number $M^*$ of channel measurements at which the throughput
is maximized decreases with the maximum allowable collision
probability $\zeta$. The reason behind this observation is that the
PM $\delta_a^*$ increases with $\zeta$ and hence less measurements
are required to achieve the same PFA (as seen from
\eqref{delta_epsilon}).

\subsubsection{Impact of Mismatched Markov Model}

\begin{figure}[!htb]
\centerline{ \scalefig{0.7}\epsfbox{ 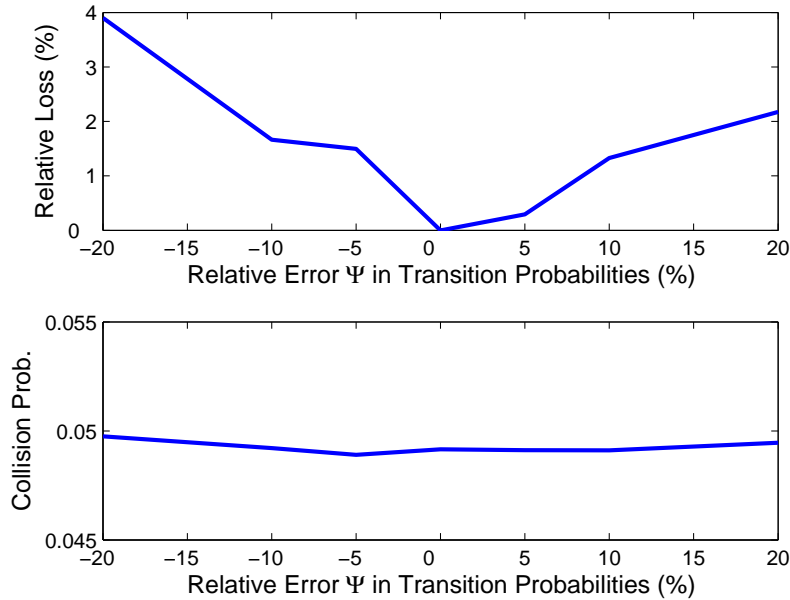}} \caption {The
impact of mismatched Markov model on the performance of the optimal
OSA strategy. } \label{fig:mc_error}
\end{figure}


We have assumed that the secondary user has perfect knowledge of the
transition probabilities of the underlying Markov model. The
transition probabilities learned by the secondary user, however, may
have errors. Suppose that the true transition probabilities are
given by $\alphabf$ and $\betabf$. The secondary user employs the
optimal OSA design based on inaccurate transition probabilities
$\alphabf'$ and $\betabf'$. In the upper half of
Fig.~\ref{fig:mc_error}, we plot the relative throughput loss as a
function of the relative estimation error $\Psi$ in transition
probabilities, where $\Psi = \frac{\alpha'_n - \alpha_n}{\alpha_n}
\times 100\% = \frac{\beta'_n - \beta_n}{\beta_n} \times 100\%$.
Note that when $\Psi = 0$, the secondary user has perfect knowledge
of the transition probabilities and hence achieves the maximum
throughput. Inaccurate knowledge can cause performance loss. We
observe that the relative throughput loss is below $4\%$ even when
the relative error is up to $20\%$. In the lower figure, we examine
the probability of collision perceived by the primary network. We
see that the probability of collision is not affected by inaccurate
transition probabilities, which confirms P1.2.


\section{OSA with Multi-Channel Sensing}\label{sec:multiple}
In this section, we address the joint design of OSA in the case
where  multiple channels can be sensed and accessed simultaneously
in each slot ($L>1$). We focus on the extension of the separation
principle developed in Section \ref{sec:one}.

\subsection{Optimal Joint Design}
Within the POMDP framework presented in Section
\ref{sec:formulation}, we first describe the three basic components
of OSA with multi-channel sensing and then derive the optimality
equation.

\subsubsection{Spectrum Sensor}
Suppose that a set $\Ac(t)\subset\{1,\ldots,N\}$ of channels is
chosen in slot $t$, where $|\Ac(t)| = L \geq 1$. The spectrum sensor
performs a $2^L$-ary hypothesis test:
\begin{equation}\label{HL}
\begin{split}
&\Hc_0 : ~\Sbf_\Ac(t) = [1,1,\ldots,1], \\
&\Hc_1 : ~\Sbf_\Ac(t) = [0,1,\ldots,1], \\
&~~~\vdots \\
&\Hc_{2^L-1}: ~\Sbf_\Ac(t) = [0,0,\ldots,0],
\end{split}
\end{equation}
where $\Sbf_\Ac(t) \,\defeq\, \{S_n(t)\}_{n\in\Ac(t)} \in \{0,1\}^L$
denotes the occupancy states of the chosen channels $\Ac(t)$ in the
current slot. The {\em a priori} probabilities of these hypotheses
can be learned from the observation and decision history, which is
characterized by the belief vector. For example, given current
belief vector $\Lambdabf(t)$ and chosen channels $\Ac(t)$, the {\em
a priori} probability of $\Hc_0$ in this slot is given by
\begin{equation}
\Pr\{\Hc_0\} = \sum_{\sbf\in\mathbb{S}}\sum_{\sbf'\in\mathbb{S}}
\lambda_{\sbf'}(t)P_{\sbf',\sbf}\prod_{n\in\Ac(t)}1_{[s_n = 1]}.
\end{equation}
This indicates that how sensor and access information at the MAC
layer can be used in the design of the spectrum sensor at the PHY
layer.

Let $\Thetabf_\Ac(t) \,\defeq\, \{\Theta_n(t)\}_{n\in\Ac(t)}
\in\{0,1\}^L$ denote the sensing outcomes. Sensing errors occur if
the spectrum sensor mistakes one hypothesis for another, \ie
$\Thetabf_\Ac(t) \neq \Sbf_\Ac(t)$. Since there are total $2^L$
hypotheses, the performance of the spectrum sensor can be specified
by a set $\Ec(t)$ of $2^L(2^L-1)$ error probabilities:
\begin{equation}
\Ec(t) \,\defeq\,\{\Pr\{\mbox{detect }\Hc_i\,|\,\Hc_j \mbox{ is
true}\}: 0\leq i, j \leq 2^L-1, i\neq j\}.
\end{equation}
The optimal design of the spectrum sensor should achieve a tradeoff
among these $2^L (2^L-1)$ error probabilities. Let
$\mathbb{A}_\delta^{(L)}(\Ac)$ include all sets of achievable error
probabilities. A sensor operating policy specifies, in each slot
$t$, a feasible sensor operating point (\ie a set of achievable
error probabilities) $\Ec(t) \in \mathbb{A}_\delta^{(L)}(\Ac(t))$
based on the current belief vector $\Lambdabf(t)$ and the chosen
channels $\Ac(t)$.

\subsubsection{Sensing and Access Policies}
At the beginning of each slot $t$, a sensing policy specifies a set
$\Ac(t)\in\mathbb{A}_s^{(L)} \,\defeq\, \{\Ac\subset \{1,\ldots,N\},
|\Ac|=L \}$ of channels to be sensed based on the current belief
vector $\Lambdabf(t)\in\Pi(\mathbb{S})$. Based on $\Lambdabf(t)$ and
the imperfect sensing outcomes $\Thetabf_\Ac(t)$ given by the
spectrum sensor, an access policy decides whether to access
$\Phibf_\Ac(t) \,\defeq\, \{\Phi_n(t)\}_{n\in\Ac(t)}\in\{0,1\}^L$.
At the end of slot $t$, the receiver acknowledges every successful
transmission. The acknowledgments (\ie the common observation of the
transmitter and the receiver) are denoted by $\Kbf_\Ac(t) \,\defeq\,
\{K_n(t)\}_{n\in\Ac(t)} \in\{0,1\}^L$, where $K_n(t) =
S_n(t)\Phi_n(t)$. Given observations $\Kbf_\Ac(t)$ and sensing
action $\Ac(t)$, the secondary user obtains an immediate reward
$R_{\Kbf_\Ac(t)}$:
\begin{equation}\label{}
R_{\Kbf_\Ac(t)} = \sum_{n\in\Ac}K_n(t) B_n.
\end{equation}

\subsubsection{Optimality Equation}
In a similar fashion as Section \ref{sec:formulation}, we can
formulate the optimal design of OSA with multi-channel sensing as a
constrained POMDP. We can also show that
Proposition~\ref{prop:deterministic} holds, \ie it is sufficient to
consider deterministic sensor operating and sensing policies for the
optimal design of OSA with multi-channel sensing. Therefore, in each
slot, the secondary user needs to make the following decisions:
which set $\Ac\in\mathbb{A}_s^{(L)}$ of channels to sense, which
sensor operating point $\Ec\in\mathbb{A}_\delta^{(L)}(\Ac)$ to
choose, and which set $\Fc \,\defeq\,
\{f_n(\thetabf)\}_{\substack{n\in\Ac\\\thetabfs\in\{0,1\}^L}}$ of
transmission probabilities to use, where
\begin{equation*}
f_n(\thetabf)= \Pr\{\Phi_n = 1\,|\,\Thetabf_\Ac = \thetabf\}
\in[0,1]
\end{equation*} is the probability of accessing chosen
channel $n$ given belief vector and sensing outcome $\Thetabf_\Ac =
\thetabf$. The composite action space is denoted by
\begin{equation*}
\mathbb{A}^{(L)} = \{\{\Ac, \Ec, \Fc\}: \Ac \in \mathbb{A}_s^{(L)},
\Ec\in\mathbb{A}_\delta^{(L)}(\Ac), \Fc \in [0,1]^{L2^L}\}.
\end{equation*}

We can obtain the optimality equation and the design constraint as
\begin{subequations}\label{V}
\begin{align}
&V_t(\Lambdabf(t)) = \max_{A = \{\Ac, \Ec, \Fc\}\in\mathbb{A}^{(L)}}
\sum_{\sbf\in\mathbb{S}} \sum_{\sbf' \in\mathbb{S}}
\lambda_{\sbf'}(t)P_{\sbf',\sbf} \sum_{\kbf_\Ac\in \{0,1\}^L}
U_{\sbf,\kbf_\Ac}^{(L)}(A) \notag\\
&\qquad\qquad\qquad\qquad\qquad\times\left[ R_{\kbf_\Ac} +
V_{t+1}(\Tc(\Lambdabf(t)\,|\,A, \kbf_\Ac))\right],
\quad 1\leq t < T, \\
&V_T(\Lambdabf(T)) = \max_{A = \{\Ac, \Ec, \Fc\}
\in\mathbb{A}^{(L)}} \sum_{\sbf\in\mathbb{S}} \sum_{\sbf'
\in\mathbb{S}} \lambda_{\sbf'}(t)P_{\sbf',\sbf}
\sum_{\kbf_\Ac\in\{0,1\}^L}
U_{\sbf,\kbf_\Ac}^{(L)}(A)  R_{\kbf_\Ac},\\
&\mbox{s.t.}~ P_n(t) = \sum_{\thetabfs_\Ac, \sbf_\Ac \in\{0,1\}^L}
h_{\Sbf_\Ac|S_n}(\sbf_\Ac\,|\,0) \,
l_{\Thetabfs_\Ac|\Sbf_\Ac}(\thetabf_\Ac\,|\,\sbf_\Ac) \,
f_n(\thetabf_\Ac) \leq \zeta,\quad \forall n, t,\label{constraintL}
\end{align}
\end{subequations}
where $ h_{\Sbf_\Ac|S_n}(\sbf_\Ac\,|\,i) \,\defeq\, \Pr\{\Sbf_\Ac =
\sbf_\Ac\,|\,S_n = i\}$ is the conditional distribution of channel
occupancy states $\Sbf_\Ac$ given current belief vector
$\Lambdabf(t)$,
$l_{\Thetabfs_\Ac|\Sbf_\Ac}(\thetabf_\Ac\,|\,\sbf_\Ac)\,\defeq\,
\Pr\{\Thetabf_\Ac = \thetabf_\Ac\,|\,\Sbf_\Ac =\sbf_\Ac\}$ is the
error probability determined by the current sensor operating point
$\Ec$, and the conditional distribution $U_{\sbf,\kbf_\Ac}^{(L)}(A)$
of observations $\Kbf_\Ac$ can be calculated as
\begin{equation}\label{UL}
\begin{split}
U_{\sbf,\kbf_\Ac}^{(L)}(A) &\,\defeq\, \Pr\{\Kbf_\Ac =
\kbf_\Ac\,|\,\Sbf = \sbf\}\\
& = \sum_{\thetabfs_\Ac\in\{0,1\}^L}
l_{\Thetabfs_\Ac|\Sbf_\Ac}(\thetabf_\Ac\,|\,\sbf_\Ac) \prod_{n\in\Ac}
\Pr\{K_n = k_n\,|\,\Thetabf_\Ac = \thetabf_\Ac, \Sbf_\Ac = \sbf_\Ac\} \\
& = \sum_{\thetabfs_\Ac\in\{0,1\}^L}
l_{\Thetabfs_\Ac|\Sbf_\Ac}(\thetabf_\Ac\,|\,\sbf_\Ac)
\prod_{n\in\Ac} [k_n s_n f_n(\thetabf_\Ac) + (1-k_n)(1 - s_n
f_n(\thetabf_\Ac))].
\end{split}
\end{equation}
The updated belief vector $\Tc(\Lambdabf(t)\,|\,A, \kbf_\Ac)$ can be
obtained by substituting \eqref{UL} into \eqref{T}.

In principle, the optimal decisions $\{\Ac^*, \Ec^*, \Fc^*\}$ in
each slot can be obtained by solving \eqref{V} recursively. However,
without any structural results on this constrained POMDP, \eqref{V}
is computationally prohibitive. A natural question here is whether
there exists a separation principle similar to Theorem \ref{th:sep}
that can be used to simplify the optimal design of OSA with
multi-channel sensing.

\subsection{Separation Principle} \label{sec:sepa}
We show that under certain conditions, the separation principle
established for the single-channel sensing case can be applied in
the multi-channel sensing scenarios.

\begin{theorem} \label{th:sep_multi}
{\em When the spectrum sensor and the access policy are designed
independently across channels, the separation principle developed in
Theorem \ref{th:sep} is valid for optimal OSA design with
multi-channel sensing. In this case, the optimal spectrum sensor
adopts the optimal NP detector with PM equal to $\zeta$, which
detects the occupancy of a chosen channel by using the measurements
from this channel, and the optimal access decision on a chosen
channel is to trust the sensing outcome from this channel. The
optimal sensing policy can be obtained by solving an unconstrained
POMDP.}
\end{theorem}

\begin{proof}
The proof is built upon that of Theorem \ref{th:sep}. See Appendix
E.
\end{proof}

We emphasize that the extension of the separation principle to
multi-channel sensing scenarios is based on the condition that the
spectrum sensor and the access policy are designed independently
across channels. Specifically, we assume that the occupancy of a
channel is detected independently of the measurements taken from
other channels and the access decision on a channel is made
independently of the sensing outcomes from other channels.
Intuitively, in this case, the design of spectrum sensor and access
policy for the multi-channel $L>1$ sensing case can be treated as
$L$ independent design problems, one for each chosen channel. Hence,
the optimal design for the single-channel case can be extended to
$L>1$.

Theorem \ref{th:sep_multi} provides sufficient conditions under
which the design given by the separation principle (referred to as
the SP approach for simplicity) is optimal. In Proposition
\ref{prop:ii}, we show that the SP approach is locally optimal (\ie
maximizes the instantaneous throughput) under certain relaxed
conditions.

\begin{proposition}\label{prop:ii}
{\em Suppose that the spectrum sensor is designed independently
across channels while the access policy jointly exploits the sensing
outcomes from all channels. The SP approach is locally optimal when
channels evolve independently.}
\end{proposition}

\begin{proof}
See Appendix F.
\end{proof}

It may sound plausible that the SP approach is (globally) optimal
when channels evolve independently since in this case the sensing
outcomes are independent across channels and independent access
decisions seem to suffice. Interestingly, counter examples can be
constructed to show that introducing correlation among access
decisions across channels can improve the overall throughput. The
rationale behind this is that the joint access design enables the
secondary user to trade the immediate access to ``bad'' channels
(\eg channels with small bandwidth) for information on the occupancy
states of ``good'' channels, leading to potentially more rewarding
future decisions. Specifically, as noted in Section \ref{sec:sepa1},
the secondary user cannot distinguish a busy channel $S_n = 0$ from
the decision of no access $\Phi_n = 0$ when observing $K_n = 0$.
However, if the access decision $\Phi_m$ on channel $m \neq n$ is
correlated with $\Phi_n$, then we can infer the occupancy state of
channel $n$ from both $K_m$ and $K_n$. That is, by sacrificing the
immediate access to channel $m$ with small bandwidth, we can obtain
more information on the occupancy state of channel $n$.

\subsection{Heuristic Approaches to Exploiting Channel Correlation}
While simplifying the design of OSA with multi-channel sensing, the
condition that the spectrum sensor and the access policy are
designed independently across channels can cause throughput
degradation since the correlation among channel occupancies is
ignored. We propose two heuristic approaches to exploit the channel
correlation: one at the PHY layer and the other at the MAC layer.

\subsubsection{Exploiting Channel Correlation at the PHY Layer}
When the occupancy states are correlated across channels, we have
correlated channel measurements at the PHY layer. Hence, the
measurements at all chosen channels should be jointly exploited in
spectrum opportunity identification. With this in mind, we propose a
heuristic design of the spectrum sensor: it performs $L$ binary
hypothesis tests, one for each chosen channel, by using all channel
measurements and adopting the optimal NP detector with PM equal to
$\zeta$. We point out that, different from the SP sensor, the
proposed spectrum sensor performs $L$ composite hypothesis tests
since it uses all channel measurements and the occupancy states of
other channels are unknown in each hypothesis test. Hence, the
structure of the optimal NP detector adopted by this heuristic
sensor relies on the joint distribution of the channel occupancy
states, which is given by the belief vector (see Section
\ref{sec:cmp} for an example). That is, the spectrum sensor design
is affected by the observation and decision history and thus varies
with time. As illustrated in Fig.~\ref{fig:ROCvsT}, the performance
of this spectrum sensor improves over time, resulting from more
informative distribution of the SOS obtained from accumulating
observations.  Note that the design of this spectrum sensor is much
simpler than the $2^L$-ary hypothesis test given in \eqref{HL}.

Based on the sensing outcomes given by this sensor that exploits
measurements from all chosen channels, access decisions are made
independently across channels, \ie access if and only if a channel
is sensed as idle. We refer this approach as the PHY layer approach.

\begin{proposition} \label{prop:ji}
{\em Suppose that the access policy is designed independently across
channels while the spectrum sensor jointly exploits the measurements
taken from all chosen channels. The PHY layer approach is locally
optimal. When channels evolve independently, the PHY layer approach
reduces to the SP approach.}
\end{proposition}

\begin{proof}
See Appendix G.
\end{proof}

Note that the PHY layer approach is locally optimal even when
channels are correlated.

\subsubsection{Exploiting Channel Correlation at the MAC Layer}
When channel occupancies are correlated, so are the sensing outcomes
given by the spectrum sensor. Hence, the channel correlation can
also be exploited at the MAC layer by making access decisions
jointly across channels. A heuristic MAC layer approach is to adopt
the spectrum sensor of the SP approach, \ie detects the occupancy
state of a channel by using only the measurements of this channel,
and then choose the access policy that exploits sensing outcomes
from all chosen channels to maximize the instantaneous throughput.
Specifically, for given chosen channels $\Ac\in\mathbb{A}_s$ and
belief vector $\Lambdabf(t)$ in slot $t$,
we choose transmission probabilities $\hat{\Fc} = \{f_n(\thetabf_\Ac)\}_{\substack{n\in\Ac\\
\thetabfs_\Ac\in\{0,1\}^L}} \in[0,1]^{L2^L}$ as follows
\begin{subequations}\label{sub1}
\begin{align}
\hat{\Fc} &= \arg\max_{\Fc \in[0,1]^{L2^L}}
\mathbb{E}\left[\left.R_{\Kbf_\Ac}\,\right|\,\Lambdabf(t)\right]\\
&= \arg\max_{\Fc \in[0,1]^{L2^L}} \sum_{n\in\Ac}B_n \Pr\{K_n = 1\}
\arg\max_{\Fc \in\mathbb{A}_c^{(L)}} \sum_{n\in\Ac}B_n
\Pr\{\Phi_nS_n = 1\} \notag\\
&= \arg\max_{\Fc \in[0,1]^{L2^L}} \sum_{n\in\Ac}B_n \Pr\{S_n = 1\}
\sum_{\thetabfs_\Ac, \sbf_\Ac \in\{0,1\}^L}
h_{\Sbf_\Ac|S_n}(\sbf_\Ac\,|\,1) \,
l_{\Thetabfs_\Ac|\Sbf_\Ac}(\thetabf_\Ac\,|\,\sbf_\Ac) \,
f_n(\thetabf_\Ac) \\
&\mbox{s.t.}~~ P_n(t) = \sum_{\thetabfs_\Ac,
\sbf_\Ac \in\{0,1\}^L} h_{\Sbf_\Ac|S_n}(\sbf_\Ac\,|\,0) \,
l_{\Thetabfs_\Ac|\Sbf_\Ac}(\thetabf_\Ac\,|\,\sbf_\Ac) \,
f_n(\thetabf_\Ac) \leq \zeta,~~\forall n \in \Ac,
\end{align}
\end{subequations}
where the conditional probability $h_{\Sbf_\Ac|S_n}(\sbf_\Ac\,|\,i)$
($i = 0,1$) of the current channel occupancies $\Sbf_\Ac$ and the
sensing error probability $l_{\Thetabfs_\Ac|\Sbf_\Ac}
(\thetabf_\Ac\,|\,\sbf_\Ac)$ are defined below \eqref{V}.

The access policy given in \eqref{sub1} can be obtained via linear
programming. Proposition \ref{prop:ij} shows that this MAC layer
approach is equivalent to the SP approach when the SOS evolves
independently across channels. This agrees with our intuition that
when channels are independent, so are the sensing outcomes from the
chosen channels. Hence, independent access decisions perform as well
as the joint one in terms of instantaneous throughput.

\begin{proposition} \label{prop:ij}
{\em Suppose that the spectrum sensor is designed independently
across channels while the access policy jointly exploits the sensing
outcomes from all chosen channels. When channels evolve
independently, the MAC layer approach reduces to the SP approach and
hence is locally optimal. }
\end{proposition}

\begin{proof}
See Appendix F.
\end{proof}

\subsection{Simulation Examples}\label{sec:cmp}
Next, we study the performance of the SP, the PHY layer, and the MAC
layer approaches. Note that these three approaches differ in the
spectrum sensor and the access policy. We can employ any sensing
policy to compare their performance. For simplicity, we consider a
myopic sensing policy that chooses the set $\Ac$ of channels to
maximize the expected instantaneous throughput under perfect
sensing: \ie for given belief vector $\Lambdabf(t)$ in slot $t$,
\begin{equation}
\Ac = \arg\max_{\Ac\in\mathbb{A}_s}\sum_{n\in\Ac} B_n \Pr\{S_n =
1\}.
\end{equation}

We adopt the model of Gaussian noise and Gaussian primary signal
described in Section \ref{sec:simulation}. In this case, the
spectrum sensor of the SP approach employs an energy detector given
in \eqref{test1}. The detection threshold $\eta_n$ of the energy
detector is chosen so that the PM is fixed at $\zeta$.

Using the measurements $\{\Ybf_n\}_{n\in\Ac}$ from all chosen
channels, the sensor employed by the PHY layer approach performs a
composite hypothesis test for each chosen channel $n$:
\begin{equation}\label{Htest2}
\begin{split}
\Hc_0 (S_n = 1): & \qquad
\Ybf_n\sim\Nc(\mathbf{0}_M,\sigma^2_{n,0}\Ibf_M), \\
& \qquad \Ybf_m \sim \Nc(\mathbf{0}_M,(\sigma^2_{m,0} + 1_{[S_m =
0]} \sigma^2_{m,1})\Ibf_M),
\quad \forall m \in\Ac\backslash\{n\}\\
\Hc_1 (S_n = 0): &  \qquad
\Ybf_n\sim\Nc(\mathbf{0}_M,(\sigma^2_{n,1} + \sigma^2_{n,0})\Ibf_M),
\\
&\qquad  \Ybf_m \sim \Nc(\mathbf{0}_M,(\sigma^2_{m,0} + 1_{[S_m =
0]} \sigma^2_{m,1})\Ibf_M), \quad \forall m \in\Ac\backslash\{n\}.
\end{split}
\end{equation}
Note that the distribution of the measurements under each hypothesis
depends on the distribution of the current channel occupancy states
$\Sbf_\Ac = \{S_n\}_{n\in\Ac}$, which is given by
$h_{\Sbf|S_n}(\sbf_\Ac\,|\,i)$ (defined below \eqref{V}) and can be
calculated from the current belief vector $\Lambdabf(t)$. In this
case, the optimal NP detector for \eqref{Htest2} is given by a
likelihood ratio test \cite[Sec.~2.5]{Trees:01book}:
\begin{equation} \label{test2}
\displaystyle \dfrac{\sum_{\sbf_\Ac\in\{0,1\}^L}
h_{\Sbf_\Ac|S_n}(\sbf_\Ac\,|\,0) \prod_{m\in\Ac} p(\Ybf_m |S_m =
s_m)}{\sum_{\sbf_\Ac\in\{0,1\}^L} h_{\Sbf_\Ac|S_n}(\sbf_\Ac\,|\,1)
\prod_{m\in\Ac} p(\Ybf_m |S_m = s_m)} \gtrless^{\Hc_1}_{\Hc_0}
\tau_n,
\end{equation}
where $h_{\Sbf|S_n}(\sbf_\Ac\,|\,0) = 0$ when $s_n \neq 0$ and
$p(\Ybf_n | S_n = s_n)$ is the PDF of independent Gaussian channel
measurements $\Ybf_n$:
\begin{equation}
p(\Ybf_n | S_n = s_n) = \prod_{i = 1}^M \frac{1}{\sqrt{2\pi
(\sigma_{n,0}^2 + 1_{[s_n = 0]} \sigma_{n,1}^2)}}
e^{-\frac{Y_{n,i}^2}{2(\sigma_{n,0}^2 + 1_{[s_n = 0]}
\sigma_{n,1}^2)}}.
\end{equation}
Note that when channel occupancies are independent, the above sensor
employed by the PHY layer approach is equivalent to that of the SP
approach, which demonstrates Proposition \ref{prop:ji}. The PFA and
the PM of this sensor can be evaluated via simulation. In each slot,
the detection threshold $\tau_n$ is chosen according to the belief
vector so that the resulting PM is fixed at $\zeta$, \ie the design
of the spectrum sensor varies with time.

As proven in Propositions \ref{prop:ii} - \ref{prop:ij}, the PHY
layer and the MAC layer approaches are equivalent to the SP approach
when channels evolve independently. We thus compare below the
performance of these three approaches in correlated channels.
Specifically, we consider $N = 4$ correlated channels, each with
bandwidth $B_n = 1$. The transition probabilities of the SOS are
given by $P_{[0000],[0111]} = 0.6$, $P_{[0000],[0000]} = 0.4$,
$P_{[0111],[0000]} = P_{[1011],[0000]} = P_{[1101],[0000]} =
P_{[1110],[0000]} = 0.2$, and $P_{[0111],[1011]} = P_{[1011],[1101]}
= P_{[1101],[1110]}= P_{[1110],[0111]} = 0.8$. The maximum allowable
probability of collision is assumed to be $\zeta = 0.05$. In each
slot, $L = 3$ channels are chosen. The spectrum sensor takes $M = 1$
measurement at each chosen channel, and the noise and the primary
signal powers are given by $\sigma_{n,0}^2 = 0$ dB and
$\sigma_{n,1}^2 = 10$ dB for all~$n$.

\subsubsection{Comparison of Sensor Performance}

\begin{figure}[htb]
\centerline{ \scalefig{0.7}\epsfbox{ 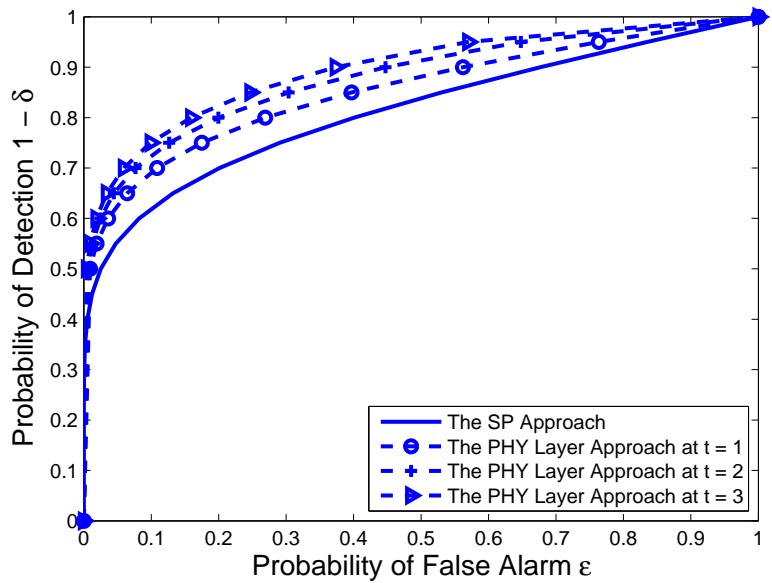}}
\caption{Comparison of ROC curves.} \label{fig:ROCvsT}
\end{figure}

In Fig. \ref{fig:ROCvsT}, we plot the ROC curves ($1 - \delta_n$ vs.
$\epsilon_n$) of the SP sensor and the sensor employed by the PHY
layer approach. Note that the sensor employed by the MAC layer
approach is the same as the SP sensor. We see that the sensor of the
PHY approach outperforms that of the SP sensor. Specifically, for a
fixed PM, the PFA of the sensor employed by the PHY approach is
smaller than that of the SP sensor. This is because the sensor of
the PHY approach exploits the correlation among channel measurements
in detection while the SP sensor uses measurements from a single
channel. We also observe that the ROC curve of the sensor of the PHY
approach improves over time while that of the SP sensor remains the
same. This observation can be explained by comparing the optimal
detectors \eqref{test1} and \eqref{test2}. Clearly, the energy
detector \eqref{test1} used by the SP sensor is static and so is its
performance. As seen from \eqref{test2}, the decision variable of
the sensor of the PHY approach depends on the conditional
distribution $h_{\Sbf_\Ac|S_n}(\sbf_\Ac\,|\,i)$ of the channel
occupancies, which varies with time according to the belief vector.
As time $t$ increases, the belief vector provides more information
on the SOS due to the accumulating observations, leading to improved
sensor performance. Fig.~\ref{fig:ROCvsT} demonstrates that the
performance of the spectrum sensor can be improved by incorporating
the sensing and access decisions at the MAC layer, which are encoded
in the belief vector.


\subsubsection{Comparison of Throughput Performance}

\begin{figure}[htb]
\centerline{ \scalefig{0.7}\epsfbox{ 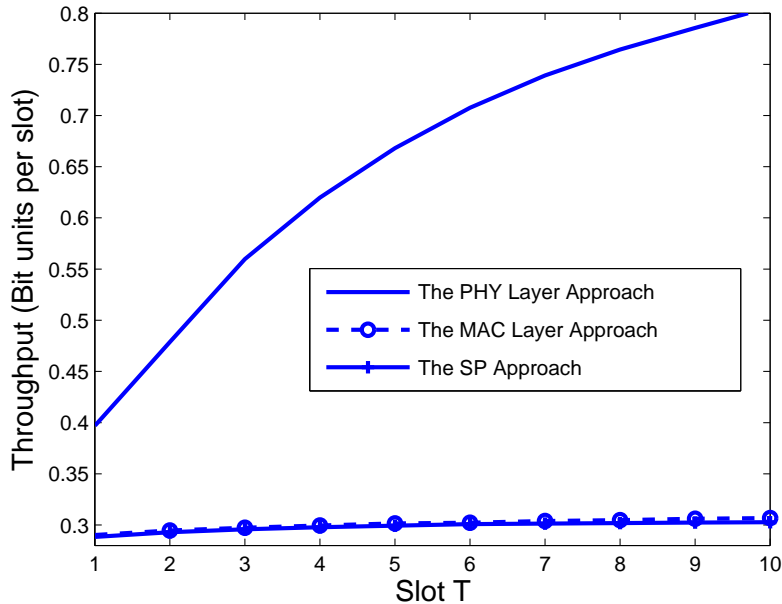}}
\caption{Comparison of normalized throughput (bit units per slot). }
\label{fig:multi_ch}
\end{figure}

In Fig. \ref{fig:multi_ch}, we compare the throughput of these three
approaches. As expected, the SP approach, which ignores the channel
correlation, performs the worst. By jointly exploiting the sensing
outcomes in access decision-making, the MAC layer approach can
improve throughput performance. A much larger performance gain is
achieved by the PHY layer approach which jointly exploits the
channel measurements in spectrum opportunity identification. We can
thus see that exploiting channel correlation at the PHY layer is
more effective than that at the MAC layer. In other words,
independent opportunity identification at the PHY layer hurts the
throughput more than independent access decision-making at the MAC
layer. This agrees with our intuition because independent
opportunity identification makes hard decisions  on whether the
channel is idle. The correlation among the resulting sensing
outcomes is less informative than that in the original channel
measurements, leading to throughput degradation.


\section{Conclusion}
Unique challenges in the design of OSA networks arise from the
tension between the secondary users' desire for performance and the
primary users' need for protection. Such tension dictates the
interaction between opportunity identification at the physical layer
and opportunity exploitation at the MAC layer, and a cross-layer
approach is necessary to achieve optimality.

In this paper, we have developed a POMDP framework that captures
basic components and design tradeoffs in OSA. We have shown that,
surprisingly, there exists a separation principle in the optimal
joint design of OSA that circumvents the curse of dimensionality in
general POMDPs. Being able to obtain the optimal joint design in
closed-form allows us to characterize quantitatively the interaction
between the physical and MAC layers. In particular, we have
demonstrated how sensing errors at the PHY layer affect MAC design
and how incorporating MAC layer information into physical layer
leads to a {\em cognitive} spectrum sensor whose performance
improves over time by learning from accumulating observations.

{\section*{Appendix A: Proof of Proposition
\ref{prop:deterministic}}

We first prove the existence of a deterministic optimal sensor
operating policy. Suppose that channel $n$ is chosen in the current
slot. Let $\omega: \mathbb{A}_\delta(n) \rightarrow[0,1]$ be an
arbitrary PDF on the set $\mathbb{A}_\delta(n)$ of feasible sensor
operating points, \ie $\int_{(\epsilon,\delta)
\in\mathbb{A}_\delta(n)}\omega(\epsilon, \delta) d\epsilon d\delta =
1$. We can compute the resulting PFA $\epsilon_n$ and the PD $1 -
\delta_n$ as
\begin{subequations}
\begin{align}
\epsilon_n &= \mathbb{E}[\epsilon] = \int_{(\epsilon,\delta)
\in\mathbb{A}_\delta(n)}\epsilon \omega(\epsilon, \delta)d\epsilon d\delta,\\
1 - \delta_n &= \mathbb{E}[1 - \delta] = \int_{(\epsilon,\delta)
\in\mathbb{A}_\delta(n)}( 1- \delta ) \omega(\epsilon,
\delta)d\epsilon d\delta.
\end{align}
\end{subequations}
Since $0\leq \epsilon \leq 1 - \delta \leq
P_{D,\max}^{(n)}(\epsilon)$ for every sensor operating point in
$\mathbb{A}_\delta(n)$, we have
\begin{equation}
0 \leq \epsilon_n \leq 1 - \delta_n \leq \int_{(\epsilon,\delta)
\in\mathbb{A}_\delta(n)}P_{D,\max}^{(n)}(\epsilon) \omega(\epsilon,
\delta)d\epsilon d\delta.
\end{equation}
Since the best ROC curve $P_{D,\max}^{(n)}$ is concave, we have
$\mathbb{E}[P_{D,\max}^{(n)}(\epsilon)] \leq
P_{D,\max}^{(n)}(\mathbb{E}[\epsilon]) $ and hence $0 \leq
\epsilon_n \leq 1 - \delta_n \leq P_{D,\max}^{(n)}(\epsilon_n)$.
That is, the resulting PFA and PM $(\epsilon_n, \delta_n) $ of any
randomized sensor operating policy $\omega$ belongs to the set
$\mathbb{A}_\delta(n)$. Therefore, it is sufficient to consider
deterministic sensor operating policies.

The spectrum sensor and the access policy should ensure that the
collision constraint is satisfied no matter which channel is chosen.
Let $v_n$ denote the maximum expected remaining reward when channel
$n$ is chosen in the current slot. Then, the deterministic sensing
policy that chooses channel $n^* = \arg\max_{n \in\mathbb{A}_s} v_n$
in this slot is optimal since the maximum expected remaining reward
that can be achieved by a randomized sensing policy is
$\sum_{n\in\in\mathbb{A}_s} v_n \mu(n) \leq v_{n^*}$, where $\mu:
\mathbb{A}_s\rightarrow [0,1]$ is a PMF on the set $\mathbb{A}_s$.

\section*{Appendix B: Proof of Theorem \ref{th:sep}}

The proof of the separation principle is built upon the following
three Lemmas. For ease of presentation, we define
$Q_t(\Lambdabf\,|\,A)$ as the maximum expected remaining reward that
can be obtained starting from slot $t$ given that the current belief
vector is $\Lambdabf$ and action $A = \{a,(\epsilon_a,
\delta_a),(f_a(0),f_a(1))\} \in \mathbb{A}$ is taken in this slot,
\ie
\begin{equation}\label{Q}
\begin{split}
Q_t(\Lambdabf\,|\,A) &= \sum_{\sbf\in\mathbb{S}} \sum_{\sbf'
\in\mathbb{S}} \lambda_{\sbf'}P_{\sbf',\sbf} \sum_{k=0}^1
U_{\sbf,k}(A) \left[ k B_a  + V_{t+1}(\Tc(\Lambdabf \,|\,A,
k))\right].
\end{split}
\end{equation}
Let $A \,\defeq\, \{a,(\epsilon_a, \delta_a),(f_a(0),f_a(1))\} \in
\mathbb{A}$ and $A' \,\defeq\, \{a,(\epsilon_a',
\delta_a'),(f_a'(0),f_a'(1))\} \in \mathbb{A}$ be two actions with
the same channel selection but different sensor operating points and
transmission probabilities.

\begin{lemma}\label{lemma:convex}
{\em The value function given in \eqref{V1} is convex in the belief
vector. Specifically, at any time $t$, the value functions
$V_t(\Lambdabf_1)$ and $V_t(\Lambdabf_2)$ of any two belief vectors
$\Lambdabf_1\in\Pi(\mathbb{S})$ and $\Lambdabf_2\in\Pi(\mathbb{S})$
satisfy
\begin{equation}\label{a.1}
V_t(\tau \Lambdabf_1 + (1-\tau)\Lambdabf_2) \leq \tau
V_t(\Lambdabf_1) + (1-\tau)V_t(\Lambdabf_2), \quad \mbox{ where
}0\leq\tau\leq 1.
\end{equation}}
\end{lemma}

\begin{proof}
We use mathematical induction. From the value function given in
\eqref{VT1}, we can see that $V_T(\Lambdabf)$ in the last slot $t =
T$ is linear and hence convex in the belief vector $\Lambdabf$.
Suppose that $V_t(\Lambdabf)$ is convex for every slot $t>t_0$. By
the definition of convex functions, we can show that the maximum
remaining reward $Q_t(\Lambdabf\,|\,A)$ under an action $A\in
\mathbb{A}$ is convex. Since the maximum of a set of convex
functions is convex, the value function $V_{t_0}(\Lambdabf)$ in slot
$t = t_0$ is convex and Lemma \ref{lemma:convex} follows.
\end{proof}

\begin{lemma}\label{lemma:=}
{\em If acknowledgement $K_a = 1$ is observed in a slot $t$, then
the future reward, given by the value function
$V_{t+1}(\Tc(\Lambdabf\,|\,A,1))$, is independent of the sensor
operating point $(\epsilon_a, \delta_a)$ and the transmission
probabilities $(f_a(0),f_a(1))$ employed in the current slot. That
is,
\begin{equation}
V_{t+1}(\Tc(\Lambdabf\,|\,A,1)) = V_{t+1}(\Tc(\Lambdabf\,|\,A',1)).
\end{equation}}
\end{lemma}

\begin{proof}
Applying the conditional observation probability $U_{\sbf,1}(A)$
given in \eqref{U1} to \eqref{T}, we obtain the updated belief
vector $\Lambdabf^1(t+1)\,\defeq\, \Tc(\Lambdabf\,|\,A,1)$ whose
element $\lambda_{\sbf}^1(t+1)$ is given by
\begin{equation}\label{1}
    \lambda_{\sbf}^1(t+1) = \frac{\sum_{\sbf'
\in\mathbb{S}} \lambda_{\sbf'}(t)P_{\sbf',\sbf} s_a} {\sum_{\sbf'
\in\mathbb{S}}\sum_{\sbf' \in\mathbb{S}}
\lambda_{\sbf'}(t)P_{\sbf',\sbf} s_a},
\end{equation}
which is independent of the sensor operating point $(\epsilon_a,
\delta_a)$ and the transmission probabilities $(f_a(0),f_a(1))$.
\end{proof}

\begin{lemma}\label{lemma:tau1}
{\em  In any slot $t$, the future rewards
$V_{t+1}(\Tc(\Lambdabf\,|\,A,k))$ and
$V_{t+1}(\Tc(\Lambdabf\,|\,A',k))$ satisfy the following inequality:
\begin{equation}\label{a.2}
\begin{split}
V_{t+1}(\Tc(\Lambdabf\,|\,A,0)) \leq \tau
V_{t+1}(\Tc(\Lambdabf\,|\,A,1)) + (1-\tau)
V_{t+1}(\Tc(\Lambdabf\,|\,A',0)),
\end{split}
\end{equation}
where $\tau$ is given by
\begin{equation}\label{tau1}
\tau = \frac{\sum_{\sbf\in\mathbb{S}}\sum_{\sbf'\in\mathbb{S}}
\lambda_{\sbf'}(t)P_{\sbf',\sbf}\left[U_{\sbf,0}(A) -
U_{\sbf,0}(A')\right] }
{\sum_{\sbf\in\mathbb{S}}\sum_{\sbf'\in\mathbb{S}}
\lambda_{\sbf'}(t)P_{\sbf',\sbf}U_{\sbf,0}(A)}.
\end{equation}}
\end{lemma}

\vspace{0.5em}

\begin{proof}
Applying the conditional observation probability $U_{\sbf,k}(A)$
given in \eqref{U1} to \eqref{T}, we can obtain the updated belief
vectors $\Tc(\Lambdabf\,|\,A,k)$ and $\Tc(\Lambdabf\,|\,A',k)$.
After some algebras, we reach the following equality:
\begin{equation}
\begin{split}
\Tc(\Lambdabf\,|\,A,0) = \tau \Tc(\Lambdabf\,|\,A,1)  + (1-\tau)
\Tc(\Lambdabf\,|\,A',0),
\end{split}
\end{equation}
where $\tau$ is given by \eqref{tau1}. Lemma \ref{lemma:tau1}
follows from the convexity of the value function proven in Lemma
\ref{lemma:convex}.
\end{proof}

With the above three Lemmas, we now prove the separation principle.
First notice that the expected immediate reward
$\mathbb{E}[R_{K_a(t)}\,|\,\Lambdabf(t)]$ can be obtained as
\begin{align}
\mathbb{E}[R_{K_a(t)}\,|\,\Lambdabf(t)] &= B_a \sum_{\sbf
\in\mathbb{S}} \sum_{\sbf' \in\mathbb{S}}
\lambda_{\sbf'}(t) P_{\sbf',\sbf} U_{\sbf,1}(A) \notag\\
&= [\epsilon_af_a(0) + (1-\epsilon_a)f_a(1)] B_a \sum_{\sbf
\in\mathbb{S}} \sum_{\sbf' \in\mathbb{S}}
\lambda_{\sbf'}(t)P_{\sbf',\sbf}s_a.
\end{align}
Since $B_a\sum_{\sbf \in\mathbb{S}} \sum_{\sbf' \in\mathbb{S}}
\lambda_{\sbf'}(t)P_{\sbf',\sbf}s_a$ is a constant for given belief
vector $\Lambdabf(t)$ and sensing action $a$, the expected immediate
reward $\mathbb{E}[R_{K_a(t)}\,|\,\Lambdabf(t)]$ increases with
$\epsilon_af_a(0) + (1-\epsilon_a)f_a(1)$.

Second, we note that the sensor operating point $(\epsilon_a,
\delta_a)$ and the transmission probabilities $(f_a(0), f_a(1))$
only affect the expected remaining reward $Q_t(\Lambdabf(t)\,|\,A)$
defined in \eqref{Q} through the observation probability
$U_{\sbf,1}(a,\delta,f(0),f(1)) = s_a [\epsilon_a f_a(0) +
(1-\epsilon_a)f_a(1)]$. Therefore, if we can show that
$Q_t(\Lambdabf(t)\,|\,A)$ increases with the quantity $\epsilon_a
f_a(0) + (1-\epsilon_a) f_a(1)$, then this will prove the separation
principle.

To this end, we consider two actions $A$ and $A'$ such that
$\epsilon_a' f_a'(0) + (1-\epsilon_a')f_a'(1) \geq \epsilon_a f_a(0)
+ (1-\epsilon_a)f_a(1)$ in slot $t$. Comparing the resulting maximum
expected remaining rewards $Q_t(\Lambdabf(t)\,|\,A')$ and
$Q_t(\Lambdabf(t)\,|\,A)$, we obtain that
\begin{align}\label{Q1}
&Q_t(\Lambdabf(t)\,|\,A') - Q_t(\Lambdabf(t)\,|\,A)\notag\\
=&\sum_{\sbf\in\mathbb{S}} \sum_{\sbf' \in\mathbb{S}}
\lambda_{\sbf'}(t)P_{\sbf',\sbf} \left\{B_a\left[U_{\sbf,1}(A') -
U_{\sbf,1}(A)\right]\right. \notag\\
&\qquad \qquad \qquad \times \sum_{k=0}^1 \left[U_{\sbf,k}(A')
V_{t+1}(\Tc(\Lambdabf(t)\,|\,A',k))- U_{\sbf,k}(A)
V_{t+1}(\Tc(\Lambdabf(t)\,|\,A,k)) \right] \} \notag\\
\geq& \sum_{\sbf\in\mathbb{S}} \sum_{\sbf' \in\mathbb{S}}
\lambda_{\sbf'}(t)P_{\sbf',\sbf} \sum_{k=0}^1 \left[U_{\sbf,k}(A')
V_{t+1}(\Tc(\Lambdabf(t)\,|\,A',k))- U_{\sbf,k}(A)
V_{t+1}(\Tc(\Lambdabf(t)\,|\,A,k)) \right]
\end{align}
Applying Lemmas \ref{lemma:=} and \ref{lemma:tau1}, we obtain after
some algebras:
\begin{equation}\label{Q2}
Q_t(\Lambdabf(t)\,|\,A') - Q_t(\Lambdabf(t)\,|\,A) \geq 0,
\end{equation}
which proves the monotonicity of the expected remaining reward
$Q_t(\Lambdabf(t)\,|\,A)$ with $\epsilon_a f_a(0) + (1-\epsilon_a)
f_a(1)$ and hence completes the proof of the separation principle.

\section*{Appendix C: Proof of Proposition \ref{prop:f}}
When $\delta_a = 1$, we have $\epsilon_a = 0$ and the objective
function $\epsilon_af_a(0) +(1-\epsilon_a)f_a(1)$ given in
\eqref{s1_obj} is maximized when $f_a^*(1) = 1$. When $\delta_a \in
[0,1)$, the constraint given in \eqref{s1} can be written as
\begin{equation}\label{con}
0\leq f_a(0)\leq \frac{\zeta - \delta_a f_a(1)}{1- \delta_a}.
\end{equation}
Applying \eqref{con} to the objective function in \eqref{s1_obj}, we
obtain that
\begin{equation}\label{G2}
\begin{split}
    \epsilon_af_a(0) +(1-\epsilon_a)f_a(1)
    \leq f_a(1) \left[1 - \frac{\epsilon_a}{1-\delta_a}\right]
    + \frac{\epsilon_a\zeta}{1-\delta_a},
\end{split}
\end{equation}
where the equality holds when $f_a(0) = \frac{\zeta - \delta_a
f_a(1)}{1- \delta_a}$. Since $1-\delta_a \geq \epsilon_a$ (see
footnote 2), the right hand side of \eqref{G2} increases with
$f_a(1)$. Hence, to maximize the objective function $\epsilon_a
f_a(0) + (1-\epsilon_a) f_a(1)$, we should choose the largest
$f_a(1)$ such that $f_a(0) = \frac{\zeta - \delta_a f_a(1)}{1-
\delta_a} \geq 0$ (see \eqref{con}). Therefore, when $\delta_a \leq
\zeta$, $f_a^*(1) = 1$ and correspondingly $f_a^*(0) = \frac{\zeta -
\delta_a}{1-\delta_a}$. When $\delta_a \geq \zeta$, $f_a^*(1) =
\frac{\zeta}{\delta_a}$ and correspondingly $f_a^*(0)= 0$.

%
%

\section*{Appendix D: Proof of Theorem \ref{th:opt}}
Applying the optimal transmission probabilities $(f_a^*(0),
f_a^*(1))$ given in Proposition \ref{prop:f} to the objective
function \eqref{s1_obj}, we obtain that
\begin{equation}\label{ga2}
\epsilon_af_a(0) +(1-\epsilon_a)f_a(1)  =
\begin{cases} 1 - \frac{\epsilon_a}
{1-\delta_a}(1-\zeta), & \delta_a \leq \zeta, \\
\frac{1-\epsilon_a}{\delta_a} \zeta, & \delta_a \geq \zeta.\\
\end{cases}
\end{equation}
Since the best ROC curve is concave \cite[Sec. 2.2]{Trees:01book},
both $\frac{\epsilon_a}{1-\delta_a}$ and
$\frac{1-\epsilon_a}{\delta_a}$ increase with $\epsilon_a$ and hence
decrease with $\delta_a$. From \eqref{ga2}, we can see that the
objective function $\epsilon_af_a(0) +(1-\epsilon_a)f_a(1)$
increases with $\delta_a$ when $\delta_a \leq \zeta$, but decreases
when $\delta_a \geq \zeta$. Hence, the maximum is achieved when
$\delta_a^* = \zeta$. Correspondingly, the optimal transmission
probabilities $(f_a^*(0), f_a^*(1))$ are given by $(0,1)$.

\section*{Appendix E: Proof of Theorem \ref{th:sep_multi}}

Let $A^{(L)} \,\defeq\, \{\Ac, \{(\epsilon_n,\delta_n)\}_{n\in\Ac},
\{(f_n(0), f_n(1))\}_{n\in\Ac} \}$ and $A_n \,\defeq\,
\{n,(\epsilon_n,\delta_n),(f_n(0), f_n(1))\}\in\mathbb{A}$, where
$A_n$ corresponds to the actions taken on chosen channel $n \in\Ac$.
When the spectrum sensor is designed independently across channels,
we can write $l_{\Thetabfs_\Ac|\Sbf_\Ac}(\thetabf_\Ac\,|\,\sbf_\Ac)
= \Pr\{\Thetabf_\Ac =\thetabf_\Ac \,|\, \Sbf_\Ac = \sbf_\Ac\} =
\prod_{n\in\Ac} \Pr\{\Theta_n =\theta_n\,|\, S_n=s_n\}$ in a product
form since the occupancy of a channel is detected independently of
the measurements at other chosen channels. When the access policy is
designed independently across channels, we have $f_n(\thetabf_\Ac) =
f_n(\theta_n)$ for all sensing outcomes $\thetabf_\Ac\in\{0,1\}^L$.
Therefore, we can write the conditional observation probability
$U_{\sbf,\kbf_\Ac}^{(L)}(A^{(L)})$ as \eqref{UL}
\begin{align}\label{ULss}
U_{\sbf,\kbf_\Ac}^{(L)}(A^{(L)}) &= \sum_{\thetabfs_\Ac\in\{0,1\}^L}
\prod_{n\in\Ac} \Pr\{\Theta_n = \theta_n\,|\,S_n = s_n\}[k_n s_n
f_n(\theta_n) + (1-k_n)(1 - s_n f_n(\theta_n))]\notag\\
& = \prod_{n\in\Ac} \sum_{\theta_n =0}^1 \Pr\{\Theta_n =
\theta_n\,|\,S_n = s_n\}[k_n s_n f_n(\theta_n) + (1-k_n)(1 - s_n
f_n(\theta_n))] \notag\\
& = \prod_{n\in\Ac} U_{\sbf,k_n}(A_n).
\end{align}
Similarly, after some algebras, the design constraint in
\eqref{constraintL} can be written as
\begin{equation}\label{constraintLss}
\begin{split}
P_n(t) 
=& \sum_{\theta_n = 0}^1 \Pr\{\Theta_n = \theta_n\,|\,S_n = 0\}
f_n(\theta_n) = (1-\delta_n)f_n(0) + \delta_nf_n(1)\leq \zeta, \quad
\forall n\in\Ac.
\end{split}
\end{equation}

Applying \eqref{ULss} to \eqref{V}, we can see that the sensor
operating point $(\epsilon_n,\delta_n)$ and transmission
probabilities $(f_n(0), f_n(1))$ of a chosen channel $n\in\Ac$
affect the maximum remaining reward only through $U_{\sbf,1}(A_n)
=s_n[\epsilon_nf_n(0) +(1-\epsilon_n)f_n(1)]$, which is independent
of the actions $\{A_m\}_{m\in\Ac\backslash\{n\}}$ taken on the other
channels. Moreover, the simplified constraint \eqref{constraintLss}
reveals that the collision probability of a channel $n$ is also
independent of the actions $\{A_m\}_{m\in\Ac\backslash\{n\}}$ taken
at other channels. Therefore, the design of the sensor operating and
access policies can be decoupled across channels. Following the same
proof as given in Appendix B, we can show that the expected
remaining reward increases with $\epsilon_nf_n(0)
+(1-\epsilon_n)f_n(1)$ of every chosen channel $n\in\Ac$.

On the other hand, the expected immediate reward
$\mathbb{E}[R_{\Kbf_\Ac(t)}\,|\,\Lambdabf(t)]$ is given by
\begin{equation}
\begin{split}
\mathbb{E}[R_{\Kbf_\Ac(t)}\,|\,\Lambdabf(t)]=&\sum_{n\in\Ac}B_n\Pr\{K_n
= 1\}  = \sum_{n\in\Ac}B_n \Pr\{S_n = 1\}[\epsilon_nf_n(0)
+(1-\epsilon_n)f_n(1)],
\end{split}
\end{equation}
which also increases with $\epsilon_nf_n(0) +(1-\epsilon_n)f_n(1)$.
Therefore, the separation principle developed in Theorem
\ref{th:sep} holds for $L >1$.

\section*{Appendix F: Proof of Propositions \ref{prop:ii} and \ref{prop:ij}}
Let $\Ac \in\mathbb{A}_s^{(L)}$ denote a set of chosen channels and
$\bar{\Ac}_n = \Ac \backslash\{n\}$ be all the set of chosen
channels excluding $n$. Since channels evolve independently, we have
$h_{\Sbf_{\bar{\Ac}_n}|S_n}(\sbf_{\bar{\Ac}_n}\,|\,0) =
h_{\Sbf_{\bar{\Ac}_n}|S_n}(\sbf_{\bar{\Ac}_n}\,|\,1)$, where
$h_{\Sbf_{\bar{\Ac}_n}|S_n}(\sbf_{\bar{\Ac}_n}\,|\,i) =
\Pr\{\Sbf_{\bar{\Ac}_n} = \sbf_{\bar{\Ac}_n}\,|\,S_n = i\}$. Hence,
given belief vector $\Lambdabf(t)$ and chosen channels $\Ac$ in slot
$t$, the myopic (\ie locally optimal) sensor operating point
$(\hat{\epsilon}_n,\hat{\delta}_n)$ and transmission probabilities
$\hat{\Fc} = \{\hat{f}_n(\thetabf_\Ac)\}$ are given by \eqref{sub1}
\begin{subequations}\label{a.f1}
\begin{align}
\{(\hat{\epsilon}_n,\hat{\delta}_n), \hat{\Fc}\}
&= \arg\max_{\substack{(\epsilon_n,\delta_n)\in\mathbb{A}_\delta \\
\Fc\in[0,1]^{L2^L}}}
\mathbb{E}\left[R_{\Kbf_\Ac(t)}\,|\,\Lambdabf(t)\right]\notag\\
&=
\arg\max_{\substack{(\epsilon_n,\delta_n)\in\mathbb{A}_\delta \\
\Fc\in[0,1]^{L2^L}}} \sum_{n\in\Ac}B_n \Pr\{S_n = 1\} \sum_{\theta_n
= 0}^1 \Pr\{\Theta_n = \theta_n\,|\,S_n = 1\} g_n(\theta_n) \notag\\
&= \arg\max_{\substack{(\epsilon_n,\delta_n)\in\mathbb{A}_\delta \\
\Fc\in[0,1]^{L2^L}}}  \sum_{n\in\Ac} B_n \Pr\{S_n = 1\}[\epsilon_ng_n(0) + (1-\epsilon_n)g_n(1)]\\
\mbox{s.t.}~~ P_n(t) &= \sum_{\theta_n = 0}^1 \Pr\{\Theta_n =
\theta_n\,|\,S_n = 0\} g_n(\theta_n)=(1-\delta_n)g_n(0) +
\delta_ng_n(1) \leq \zeta ,~~\forall n \in\Ac,
\end{align}
\end{subequations}
where $g(\theta_n) \in [0,1]$ is defined as
\begin{equation}\label{gn}
\begin{split}
g_n(\theta_n) \,\defeq\, & \sum_{\thetabfs_{\bar{\Ac}_n}
\in\{0,1\}^{L-1}} f_n(\thetabf_{\bar{\Ac}_n}, \theta_n)
\sum_{\sbf_{\bar{\Ac}_n} \in\{0,1\}^{L-1}} \Pr\{\Sbf_{\bar{\Ac}_n} =
\sbf_{\bar{\Ac}_n}\} \prod_{m\in\bar{\Ac}_n} \Pr\{\Theta_m =
\theta_m\,|\,S_m = s_m\}.
\end{split}
\end{equation}
We see from \eqref{a.f1} that the myopic approach should maximize
$\epsilon_ng_n(0) + (1-\epsilon_n)g_n(1)$ under the constraint
$(1-\delta_n)g_n(0) + \delta_ng_n(1)\leq \zeta$ for every chosen
channel $n\in\Ac$, leading to the same optimization problem as
\eqref{s1}. By Theorem \ref{th:opt}, $\hat{\delta}_n = \zeta$ and
$(\hat{g}_n(0), \hat{g}_n(1)) = (0,1)$ are the solution to
\eqref{a.f1}. That is, the SP sensor is locally optimal.
Furthermore, since $(\hat{g}_n(0), \hat{g}_n(1)) = (0,1)$ is
achieved by choosing $\hat{f}_n(\thetabf_{\bar{\Ac}_n}, \theta_n) =
1_{[\theta_n=1]}$ in \eqref{gn}, transmission probabilities
$\hat{f}_n(\thetabf_\Ac) = \theta_n$ are locally optimal, which
completes the proof of Proposition \ref{prop:ii}.

Proposition \ref{prop:ij} follows directly from the fact that the
MAC layer approach employs the myopic access policy and the SP
sensor, which has been proven to be locally optimal.

\section*{Appendix G: Proof of Proposition \ref{prop:ji}}
When the access policy is designed independently across channels, we
have $f_n(\thetabf_\Ac) = f_n(\theta_n)$ for any sensing outcome
$\Thetabf_\Ac = \thetabf_\Ac$ from chosen channels $\Ac$. Hence,
given belief vector $\Lambdabf(t)$ and chosen channels $\Ac$ in slot
$t$, the myopic spectrum sensor $\hat{\Ec}$ and access decisions
$\{(\hat{f}_n(0),\hat{f}_n(1))\}_{n\in\Ac}$ are given by
\begin{subequations}\label{a.g}
\begin{align}
&\{\hat{\Ec}, \{(\hat{f}_n(0),\hat{f}_n(1))\}_{n\in\Ac}\} =
\arg\max_{\substack{\Ec\in\mathbb{A}_\delta^{(L)}\\
f_n(0), f_n(1) \in[0,1]}} \sum_{n\in\Ac} B_n \Pr\{S_n =
1\}[\Pr\{\Theta_n = 1\,|\, S_n = 1\}f_n(1)\notag
\\
&\qquad\qquad\qquad \qquad\qquad\qquad\qquad
\qquad\qquad\qquad\qquad +
\Pr\{\Theta_n = 0 \,|\, S_n = 1\} f_n(0)]\\
&\mbox{s.t.}~~ P_n(t) = \Pr\{\Theta_n = 1\,|\,S_n = 0\}f_n(1) +
\Pr\{\Theta_n = 0 \,|\, S_n = 0\} f_n(0) \leq \zeta, ~~\forall n
\in\Ac,
\end{align}
\end{subequations}
where \begin{equation} \Pr\{\Theta_n = \theta_n \,|\, S_n = s_n\} =
\sum_{\thetabfs_{\bar{\Ac}_n}, \sbf_{\bar{\Ac}_n}\in\{0,1\}^{L-1}}
\Pr\{\Thetabf_{\bar{\Ac}_n} = \thetabf_{\bar{\Ac}_n}, \Theta_n =
\theta_n\,| \Sbf_{\bar{\Ac}_n} = \sbf_{\bar{\Ac}_n}, S_n = s_n\}
\end{equation} is determined by the sensor operating point
$\Ec\in\mathbb{A}_\delta^{(L)}$. Since \eqref{a.g} has the same form
as \eqref{s1}, the PHY layer approach is locally optimal.

Furthermore, when the SOS evolves independently across channels, the
measurements from different channels are independent. Hence, the
sensor employed by the PHY layer approach is equivalent to the SP
sensor.

}

\bibliographystyle{ieeetr}

\end{document}